\documentclass[aps,prd,twocolumn,eqsecnum,amssymb,amsmath,showpacs,a4paper,superscriptaddress,nofootinbib,longbibliography]{revtex4-2}
\usepackage{graphicx}
\usepackage{amsfonts}
\usepackage{amsmath}
\usepackage{hyperref}

\begin{document}

\title{Quasinormal modes, quasibound states, scalar clouds, and superradiant instabilities of a Kerr-like black hole}

\author{Pedro Henrique Croti Siqueira}
 \email{pedro.croti@ufabc.edu.br}
 \affiliation{Centro de Ci\^encias Naturais e Humanas, Universidade Federal do ABC (UFABC), 09210-170 Santo Andr\'e, S\~ao Paulo, Brazil}
\author{Maur\'icio Richartz}
 \email{mauricio.richartz@ufabc.edu.br}
\affiliation{Centro de Matem\'atica, Computa\c c\~ao e Cogni\c c\~ao, Universidade Federal do ABC (UFABC), 09210-170 Santo Andr\'e, S\~ao Paulo, Brazil}

\begin{abstract}
  We use the continued fraction method to determine the eigenfrequencies associated with a scalar field around a Kerr-like black hole. The Kerr-like metric considered in this article is a subclass of the general parametrization of axisymmetric black holes proposed by Konoplya, Rezzolla and Zhidenko. In addition to its mass $M$ and specific angular momentum $a$, the black hole depends on a third parameter $\eta$, called the deformation parameter. We investigate how the deformation parameter affects the quasinormal modes and the quasibound states of a massive scalar field around the black hole. In particular, we compute the time scales associated with the superradiant instabilities of the scalar field in such a spacetime. The properties of stationary scalar clouds that could be formed by these instabilities are also discussed. 
\end{abstract}

\maketitle

\section{Introduction}
The era of gravitational wave (GW) astronomy, initiated by the detection of the binary black hole merger GW150914~\cite{gw150914}, opened up a new avenue to test both the weak and the strong regimes of gravity. When two compact objects merge, the gravitational wave signal produced can be roughly divided into three distinct parts. The first is the inspiral phase, in which the objects approach each other. The second is the catastrophic merger, when the two objects become one, and from which most of the gravitational wave energy is emitted. The third is the ringdown phase, characterized by the emission of quasinormal modes (QNMs), in which the final object relaxes towards equilibrium~\cite{Nollert;1999,Berti;2009,Zhidenko;2011}. If the final object is a Kerr black hole, its QNM spectrum is fully characterized by only two parameters: its mass $M$ and its specific angular momentum $a$. 

Through the analysis of the ringdown signal detected by GW observatories, one can try to  identify the parameters $M$ and $a$ in order to test the Kerr hypothesis, i.e.~the conjecture that isolated astrophysical black holes are characterized by the Kerr metric~\cite{Dreyer:2003bv,Berti:2005ys,Berti:2007zu,Berti:2016lat,Brito:2018rfr,Ota:2019bzl,LIGOScientific:2020tif}.       
In fact, there are many techniques used to measure the spin of a black hole and to test the Kerr hypothesis~\cite{Bambi:2011mj}. Unfortunately, the uncertainties in the current astronomical data leave the door open for alternative theories of gravity and non-Kerr descriptions of astrophysical black holes. In particular, it allows for the possibility that the ringdown signals observed in GW detectors so far is associated with a non-Kerr black hole (see Refs.~\cite{Berti:2018vdi,Finch:2021qph,Cotesta:2022pci,Isi:2022mhy} for recent discussions on the difficulties of measuring the QNM spectrum). 

A robust and general parametrization that can be used to describe non-Kerr black holes is the one introduced by Konoplya, Rezzolla and Zhidenko (KRZ) in Ref.~\cite{KRZ;2016}. It has been recently used to constrain deviations from the Kerr hypothesis with the iron-line
method~\cite{Ni:2016uik,Cardenas-Avendano:2016zml,Nampalliwar:2019iti}. A particular subclass of the KRZ parametrization, whose deviations from the Kerr metric are described by a single deformation parameter, is the metric introduced in Refs.~\cite{Caspar:2012ux,Schonenbach:2013nya} in the context of pseudo-complex General Relativity. Using such a metric, Ref.~\cite{KZ;2016} demonstrated, through the WKB method, that non-negligible deviations from Kerr can produce QNMs which are compatible with the QNMs of a Kerr black hole. The same one-parameter subclass of deviations from Kerr was studied in Ref.~\cite{Stefano2021}, where a complete characterization of the spacetime structure and a detailed analysis of superradiant scattering were performed. 

The phenomenon of superradiance, according to which wave amplification can occur upon reflection by a rotating object like a black hole~\cite{Brito:2015oca,zeldovich1,zeldovich2,misner,staro1,staro2,Richartz:2009mi,Torres:2016iee}, can also trigger instabilities when the perturbing field is massive~\cite{Press:1972zz,Detweiler:1980uk,Furuhashi:2004jk,Dolan2007,Dolan:2012yt}. Such superradiant instabilities are closely related to the existence of quasibound states (QBSs) and may have important astrophysical implications~\cite{Brito:2015oca}. For instance, their evolution leads to the growth of a bosonic cloud around the black hole~\cite{Hod:2012px,Herdeiro:2014goa,Benone:2014ssa,Brito:2014wla,East:2017ovw,Herdeiro:2017phl}. Additionally, the detection of gravitational waves emitted by these clouds would be an evidence of the existence of ultralight bosons, with possible consequences to the physics beyond the Standard Model. The lack of such signals so far has been used to constrain the mass of ultralight fields in nature~\cite{Arvanitaki:2009fg,Arvanitaki:2010sy,Pani:2012vp,Brito:2013wya}. In Ref.~\cite{Stefano2021}, the possibility of superradiant instabilities in a deformed Kerr black hole was briefly investigated using approximate analytical methods.   

In this work, we employ the continued fraction method to determine the QNMs and the QBSs of a massive scalar field in the deformed Kerr black hole considered in Refs.~\cite{KZ;2016,Stefano2021}. Our goal is twofold. First, motivated by the possibility of testing the Kerr hypothesis though black hole spectroscopy, we compare the QNMs and QBSs of a Kerr black hole with the QNMs and QBSs of the deformed Kerr black hole. Second, inspired by the possibility of using rotating black holes as astrophysical particle detectors, we determine how the mass of scalar clouds and how the  superradiant instability timescales change according to the Kerr deformation parameter.

Our paper is organized as follows. In Sec.~II we review the propagation of a massive scalar field in a Kerr-like spacetime described by the subclass of the KRZ metric studied in Refs.~\cite{KZ;2016,Stefano2021}. In Sec.~III we set up the continued fraction method to find the eigenvalues of the associated wave equation. Our main results are presented in Sec.~IV, where we investigate the dependence of the QNMs and the QBSs on the Kerr deformation parameter. In particular, we study the properties of scalar clouds and perform a detailed analysis of the superradiant instability timescales. The last section (Sec.~V) is dedicated to our final remarks. Throughout this work we use $G = c = \hbar = 1$ units.


\section{The Klein-Gordon field in the deformed-Kerr background} 
\label{Sec:def_Kerr}
A class of asymptotically flat, stationary and axisymmetric spacetimes which possess Kerr-like symmetries and admit a separable Klein-Gordon  equation is characterized by the line element~\cite{Konoplya:2018arm}
\begin{align}
	ds^{2}=-\frac{N^{2}-W^{2}\sin ^{2}\theta }{K^{2}}dt^{2}-2Wr\sin ^{2}\theta dtd\phi \nonumber \\ +K^{2}r^{2}\sin ^{2}\theta \, d\phi ^{2}+\Sigma \left( \frac{B^{2}}{N^{2}}dr^{2}+r^{2}d\theta ^{2}\right),	\label{eq:full_deformed_Kerr_line_element}
\end{align}
with
\begin{align}
	B &= B\left( r,\theta \right) =  R_{B}(r),  \\ 
	\Sigma &= \Sigma\left( r,\theta \right) = R_{\Sigma}(r)+\frac{a^{2} \cos^{2}\theta}{r^2}, \\
	N^{2} &= N^{2}\left( r,\theta \right) = R_{\Sigma}(r) - \frac{R_{M}(r)}{r} + \frac{a^{2}}{r^{2}}, \\
	W  &= W\left(r,\theta \right) = \frac{a R_{M}(r)}{r^2 \Sigma(r, \theta)}, \\
	K^{2} &= K^{2}\left( r,\theta \right) = \frac{1}{\Sigma (r, \theta)}\left[R^{2}_{\Sigma}(r) + \frac{a^{2}}{r^2} R_{\Sigma}(r) \right. \nonumber \\ &+ \left. \frac{a^{2}\cos^{2}\theta}{r^2}  N^{2}(r,\theta)\right] + \frac{a W(r, \theta)}{r}.
	 \end{align}
We note that the functions $B$, $\Sigma$, $N^2$, $W$, and $K^2$ depend on the constant parameter $a$ and upon three arbitrary functions of the radial coordinate, namely $ R_{B}$, $ R_{\Sigma}$ and $ R_{M}$. Asymptotic flatness implies that $ R_{M} = 2M + \mathcal{O}(1/r)$, where $M$ is a constant. By redefining the radial variable, one can conveniently fix $ R_{B}$ or $ R_{\Sigma}$. To recover the Kerr spacetime, one has to choose $R_{\Sigma} = R_{B} = 1$ and $ R_{M} = 2M$. The Kerr-Newman metric, on the other hand, corresponds to $R_{\Sigma} = R_{B} = 1 $ and $ R_{M} = 2M-Q^2/r$, while the Kerr-Sen metric is obtained with $R_{\Sigma} = 1 + 2b/r$, $R_{B} = 1$ and $ R_{M} = 2M$.   

The deformed Kerr background that we consider in this work is the subclass of the metric \eqref{eq:full_deformed_Kerr_line_element}
obtained by setting $ R_{\Sigma} = R_{B} = 1 $ and $ R_{M} = 2M + \eta/r^{2}$~\cite{Caspar:2012ux,Schonenbach:2013nya,KZ;2016,Stefano2021,Konoplya:2018arm}. With this choice, the line element \eqref{eq:full_deformed_Kerr_line_element} becomes 
\begin{align}
  ds^{2}=-\frac{\Delta-\widetilde{W}^{2} r^2 \, \sin ^{2}\theta }{\widetilde{K}^{2}} \, dt^{2}-2\widetilde{W}r\sin ^{2}\theta \, dt \,d\phi \nonumber \\ +\widetilde{K}^{2}\sin ^{2}\theta \, d\phi ^{2}+ \frac{\widetilde{\Sigma}}{\Delta} \,  dr^{2} + \widetilde{\Sigma} \, d\theta ^{2},
  \label{eq:deformed_Kerr_line_element}
\end{align}
with
\begin{align}
 \Delta &= r^2 N^2 =r^{2}-2Mr +a^{2} -\frac{\eta }{r}, \label{deltafun} \\ 
\widetilde{\Sigma} &= r^2\Sigma = r^{2}+a^{2}\cos ^{2}\theta, \\
\widetilde{K}^{2} &= r^2 K^{2} =\frac{\left( r^{2}+a^{2}\right) ^{2}-a^{2}\Delta\sin ^{2}\theta  }{r^{2}+a^{2}\cos ^{2}\theta}, \\ 
 \widetilde{W}  &= W =\frac{a}{r^{2}+a^{2}\cos ^{2}\theta}\left(2M + \frac{\eta}{r^2}\right). \end{align}
We note that our notation for the  metric coefficients in \eqref{eq:deformed_Kerr_line_element} is slightly different than the notation employed in Refs.~\cite{KZ;2016,Stefano2021,Konoplya:2018arm}. 
The parameters $M$ and $a$ represent, respectively, the mass and the specific angular momentum of the spacetime. The parameter $\eta$ is the deformation parameter that quantifies deviations from the Kerr metric. When $\eta = 0$, Eq.~\eqref{eq:deformed_Kerr_line_element} reduces to the usual Kerr line element in Boyer-Lindquist coordinates. 

As is the case for the Kerr metric, depending on the parameters chosen, the deformed Kerr line element \eqref{eq:deformed_Kerr_line_element}  may represent a black hole or a naked singularity. In this work we consider only black holes, meaning that at least one of the three roots of \eqref{deltafun} is real and positive~\cite{Stefano2021}.
The location of the event horizon corresponds to the largest positive root of $\Delta$, which we denote by $r_0$. The two other roots of the $\Delta$ function are denoted by $r_1$ and $r_2$. If $r_1$ and $r_2$ are both real, we set $r_1 \ge r_2$. On the other hand, if they are complex conjugates, we set $\mathrm{Im}(r_1)>\mathrm{Im}(r_2)$.    
These roots are related to the spacetime parameters through the following identities:
\begin{equation}
\begin{cases}
r_0+r_1+r_2 = 2M, \\
r_0r_1 + r_0r_2 + r_1r_2 = a^2, \\
r_0r_1r_2 = \eta.
\end{cases}
\end{equation}

\begin{figure}[!htbp]
\centering
  \includegraphics[width = 0.95 \linewidth]{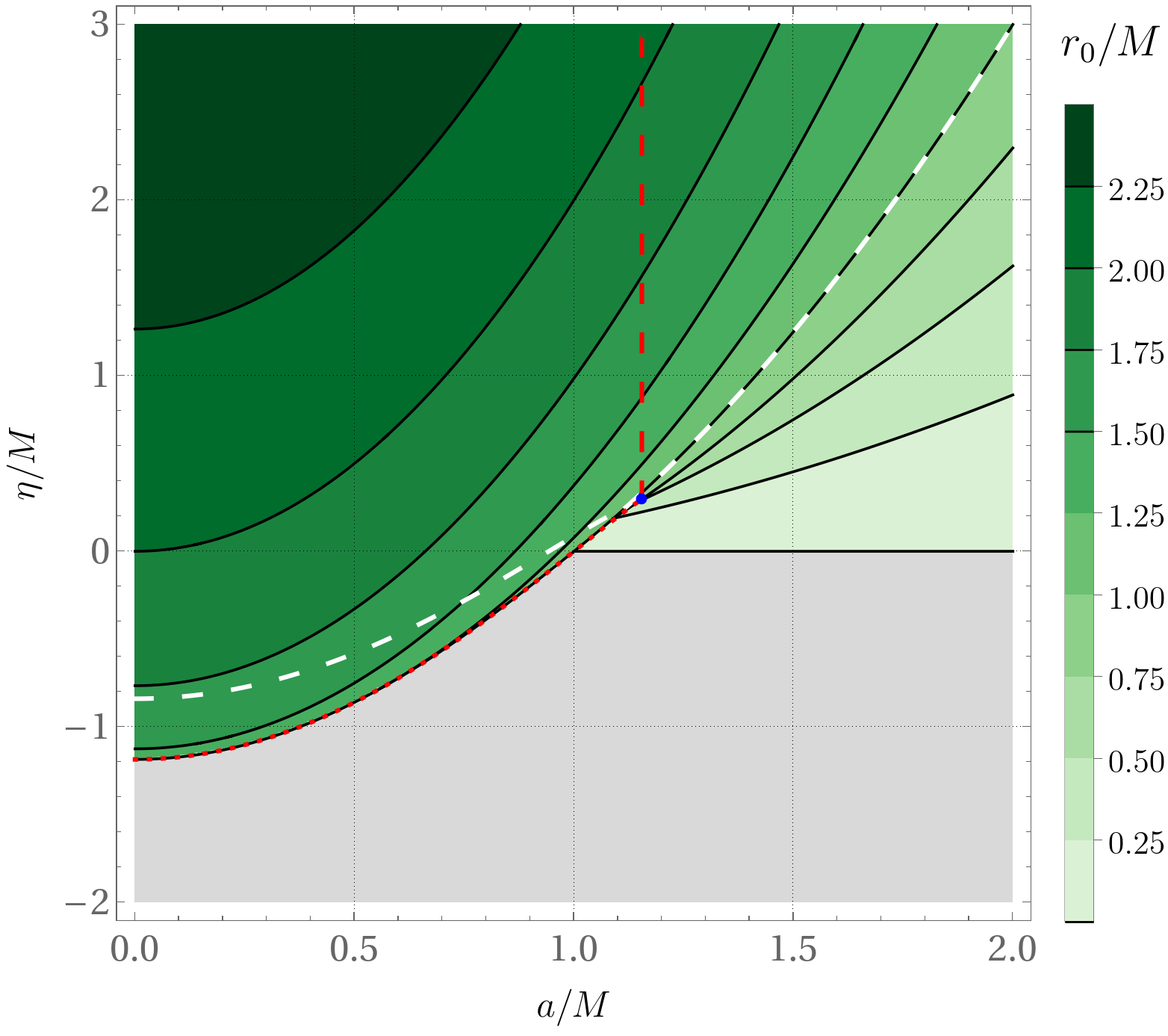}
  \caption{The location $r_0/M$ of the event horizon of the deformed Kerr black hole \eqref{eq:deformed_Kerr_line_element} as a function of the angular momentum $a/M$ and the deformation parameter $\eta/M$. For parameters in the gray region, the metric represents a naked singularity instead of a black hole. The red dotted curve and the red dashed line determine the region where the continued fraction method associated with \eqref{eq:radial_solution_KZ} converges, while the white dashed curve determines the region where the continued fraction method associated with \eqref{eq:radial_solution_KZ2} converges. Along the red dotted line the event horizon is a double root of \eqref{deltafun}. At the blue point the event horizon is a triple root.}
  \label{fig:horizons} 
\end{figure}

 In Fig.~\ref{fig:horizons} we plot the location of the event horizon along the parameter space $\left(a/M, \eta/M\right)$.  We remark that $\Delta$ admits repeated roots if
\begin{equation}
\eta = \eta_{\pm} = \frac{2}{27}\left[9Ma^{2}- 8M^{3}\pm (4M^{2}-3a^{2})^{3/2}\right].
\end{equation}
The event horizon is a double root only when $\eta=\eta_-$. In this work, we refer to the black hole as extreme whenever this condition is satisfied. The red dotted curve in Fig.~\ref{fig:horizons} identifies the extremal black holes in the parameter space. The blue point at $(2/\sqrt{3},8/27)$ corresponds to the only pair of parameters for which the event horizon is a triple root of $\Delta$. We shall refer to it as the triple point.

The deformed Kerr black hole \eqref{eq:deformed_Kerr_line_element} preserves several interesting features of the Kerr black hole. Besides the spherical symmetry of the event horizon and the same post-Newtonian asymptotics, it allows the separation of variables in the Hamilton-Jacobi equation, in the Klein-Gordon (KG) equation, and in Maxwell's equations~\cite{KZ;2016,Konoplya:2018arm,Stefano2021}. In the case of a scalar field $\Phi$ of mass $\mu$ satisfying the KG equation
\begin{equation} \label{eq:KG}
	\dfrac{1}{\sqrt{-g}}\partial _{\mu }\left(  g^{\mu \nu }\sqrt{-g}\partial _{\nu }\Phi \right) +\mu ^{2}\Phi =0,  
\end{equation}
separation of variables follows from the mode \emph{ansatz}
\begin{equation}
 \Phi = \Phi_{\omega \ell} (t, \phi, \theta, r) = e^{-i\omega t + im\phi} R(r) S(\theta),
 \label{eq:ansatz}
\end{equation}
where $\omega$ is the frequency and $m \in \mathbb{Z}$ is the azimuthal number. 

After substituting \eqref{eq:ansatz} into \eqref{eq:KG}, we find that the functions $S$ and $R$ satisfy, respectively,
\begin{align}
	\frac{1}{\sin \theta} \frac{d}{d \theta} \left( \sin \theta \frac{dS}{d \theta} \right)  +\bigg(\! \! -a^{2}q^2\cos ^{2}\theta  +\lambda -  \dfrac{m^{2}}{\sin ^{2}\theta }\bigg) S =0 \label{eq:angular_eq}
\end{align} 
and
 \begin{align} \label{eq:radial_eq}
 	\Delta \frac{d}{dr}\left(\Delta \frac{dR}{dr} \right)  +\bigg[ \omega ^{2}\left( r^{2}+a^{2}\right) -4Mam\omega r \nonumber \\ +m^{2}a^{2} -\left(\omega ^{2}a^{2}+\mu ^{2}r^{2}+ \lambda \right) \Delta \bigg] R =0,
 \end{align} 
where $q^2 = \mu ^{2} - \omega ^{2}$ and $\lambda$ is a separation constant.
As in the case of the Kerr black hole, the angular equation~\eqref{eq:angular_eq} is a spheroidal wave equation~\cite{Berti2005}. Only a discrete set of separation constants $\lambda=\lambda_{\ell m}(q)$, indexed by the positive integer $\ell \ge |m|$, yields solutions that are regular at the poles $\theta = 0$ and $\theta = \pi$. Such solutions are denominated spheroidal harmonics. Several methods~\cite{Leaver1985,Hughes:1999bq,2003JPhA...36.5477F,Cook:2014cta} can be employed to determine the eigenvalues $\lambda_{\ell m}$. 

The radial equation has also the same form of the corresponding equation for the Kerr metric, the only difference being the function $\Delta$ that depends on the deformation parameter $\eta$. Taking into account the fact that perturbations cannot escape from inside the black hole, we must impose a purely ingoing boundary condition at the event horizon, i.e.
 \begin{equation} \label{eq:BC_horizon}
	\lim _{r\rightarrow r_{0}}R\left( r\right) \sim \left( r-r_{0}\right) ^{-i \sigma_{0}},
	\end{equation}
where 
\begin{align}  \label{sigma0}
\sigma_{0} &= \frac{r_{0}(r_{0}+r_{1})(r_{0}+r_{2})}{(r_{0}-r_{1})(r_{0}-r_{2})}\left( \omega - m \Omega_0 \right).
\end{align}
The parameter $\Omega_0$, given by
\begin{equation} \label{Omega0}
\Omega_0 = \frac{a}{(r_0+r_1)(r_0+r_2)} = \frac{a}{r_0^2 + a^2},  
\end{equation} 
represents the angular velocity of the black hole. Analogous expressions for $\sigma_1$ and $\sigma_2$, which will be used in the next section, are defined by interchanging the corresponding indices in Eq.~\eqref{sigma0}. We also note that the parameters $\sigma_i$ satisfy the following identity:
\begin{equation} 
\sigma_0 + \sigma_1 + \sigma_2 = (r_0+r_1+r_2) \omega = 2M\omega.
\end{equation}

QNMs and QBSs are \emph{eigensolutions} of the wave equation that correspond to specific boundary conditions far away from the black hole. A unified description of such solutions is possible by requiring the following asymptotic behaviour of $R(r)$, which is compatible with Eq.~\eqref{eq:radial_eq}:
 \begin{equation} \label{eq:BC_infinity}
 	\lim _{r\rightarrow \infty}R\left( r\right) \sim e^{qr}r^{-1+\chi },
 	\end{equation}
where
\begin{equation}
	\chi =\frac{(r_{0}+r_{1}+r_{2})(q^{2} -  \omega ^{2})}{2q} = \frac{M(q^{2} -  \omega ^{2})}{q}.
\end{equation}
QBSs decay exponentially far away from the black hole ($\mathrm{Re}(q) < 0$), while QNMs are purely outgoing far way from the black hole ($\mathrm{Im}(q) > 0$)~\cite{Dolan2007}. The associated \emph{eigenfrequencies} $\omega$ are complex numbers whose real part $\mathrm{Re}(\omega)$ is the oscillation frequency and whose imaginary part $\mathrm{Im}(\omega) <0$ determines the decay timescale $\tau = |\mathrm{Im}(\omega)|^{-1}$. 

When the field is massive and superradiance is possible, unstable modes characterized by $\mathrm{Im}(\omega) > 0$ arise~\cite{Press:1972zz,Detweiler:1980uk,Furuhashi:2004jk,Dolan2007,Dolan:2012yt}. The oscillation frequency of such modes typically satisfies the condition for superradiance, which in the present case reads
\begin{equation}
0 < \mathrm{Re}(\omega) < m \Omega_0.
\end{equation}
For unstable modes the quantity $\tau$ represents the timescale associated with the growth of the instability.

Scalar clouds~\cite{Hod:2012px,Herdeiro:2014goa,Benone:2014ssa}, on the other hand, are bound states characterized by $\mathrm{Im}(\omega) = 0$ and, therefore, do not decay nor grow in time. Given the presence of the event horizon, such behavior is only possible if there is no energy flux towards the black hole. This requirement implies that the frequency $\omega_{\mathrm sc}$ of the scalar cloud must be synchronized with the rotation of the black hole, i.e.~
\begin{equation} \label{scalarcloudsdef}
\omega_{\mathrm sc} = m \Omega_0.
\end{equation}
Besides the synchronized frequency, the scalar cloud is also characterized by the associated mass $\mu_{\mathrm sc}$ which guarantees that the asymptotic boundary condition \eqref{eq:BC_infinity} is satisfied.


\section{Continued Fraction Method}
The numerical method we employ to determine the eigenfrequencies of the deformed Kerr black hole \eqref{eq:deformed_Kerr_line_element} is the continued fraction method~\cite{Leaver1985}. To implement the method, we first note that the differential equation \eqref{eq:radial_eq} has four regular singular points ($0$,$r_0$,$r_1$,$r_2$) and one irregular singular point ($\infty$). Taking \eqref{eq:BC_horizon} and \eqref{eq:BC_infinity} into account, we introduce the following \emph{ansatz} 
\begin{align}
	R\left( r\right)&=e^{qr}\left( r-r_{0}\right) ^{-i \sigma_{0}}  (r-r_{2})^{-i \sigma_{2}}  \nonumber \\ \times & \left( r-r_{1}\right)^{-i\sigma_1 -1+\chi + 2iM\omega} \sum ^{\infty }_{n=0}a_{n}\left( \frac{r-r_{0}}{r-r_{1}}\right) ^{n}, \label{eq:radial_solution_KZ}
\end{align}   
where $a_n \in \mathbb{C}$.
Substituting the series above into \eqref{eq:radial_eq}, we get a five-term recurrence relation for the series coefficients $a_{n}$ (the last equation holds for $n \geq 3$):
\begin{align} \label{eq:5-terms} \begin{cases} \alpha _{1}a_{2} + \beta _{1}a_{1} + \gamma _{1}a_{0} =0, \\ \alpha _{2}a_{3} + \beta _{2}a_{2} + \gamma _{2}a_{1} + \delta _{2}a_{0}=0,  \\ \alpha _{n}a_{n+1} + \beta _{n}a_{n} + \gamma _{n}a_{n-1} + \delta _{n}a_{n-2} + \epsilon _{n}a_{n-3} =0, \end{cases}
\end{align}
where $\alpha_{n}$, $\beta_{n}$, $\gamma_{n}$, $\delta_{n}$ and $\epsilon_{n}$ are functions of the parameters that characterize the black hole and the scalar field. After a double gaussian elimination~\cite{Leaver1990}, this five-term recurrence relation reduces to a 3-term recurrence relation 
\begin{equation}\label{eq:3-terms}
	\alpha^{''} _{n}a_{n+1} + \beta^{''} _{n}a_{n} + \gamma^{''} _{n}a_{n-1} =0, \qquad n \geq 1,
\end{equation}
where the new coefficients $\alpha^{''}_{n}$, $\beta^{''}_{n}$, $\gamma^{''}_{n}$ are also functions of the parameters that characterize the black hole and the scalar field.

The \emph{ansatz} \eqref{eq:radial_solution_KZ} is associated with the M\"obius transformation $z = (r-r_0)/(r-r_1)$,
which takes the event horizon, $r=r_0$, and the point at infinity, $r=\infty$, respectively into $z=0$ and $z=1$. The other three singular points, namely $r=0$, $r=r_1$ and $r=r_2$, are taken, respectively, into
\begin{equation}
z=z_0=\frac{r_0}{r_1}, \quad z=\infty, \quad z=z_2 = \frac{r_2-r_0}{r_2-r_1}.
\end{equation}
If $|z_0|>1$ and $|z_2|>1$, there are no singular points other than the event horizon inside the unit circle of the $z$-plane. The corresponding region of the $\left(a/M, \eta/M\right)$ parameter space where this happens is delimited by the red curves in Fig.~\ref{fig:horizons} (note that the red dotted curve also identifies the extremal black holes). Inside this region, the Frobenius method guarantees that the series in \eqref{eq:radial_solution_KZ} converges everywhere in the interval $ r_0 \le r<\infty$. Convergence at $r=\infty$ occurs if and only if the following continued fraction equation is satisfied:
\begin{eqnarray} \label{cfequ}
	0=\beta_{0}^{''}-\frac{\alpha_{0}^{''} \gamma_{1}^{''}}{\beta_{1}^{''}-}\frac{\alpha_{1}^{''} \gamma_{2}^{''}}{\beta_{2}^{''}-}\frac{\alpha_{2}^{''} \gamma_{3}^{''}}{\beta_{3}^{''}-} \ldots \text{\space}.
\end{eqnarray}

For practical purposes, the continued fraction above must be truncated at some order $n$ and a root finding algorithm must be used to determine the solutions of the equation. The rate of convergence of the method (in terms of $n$) can be improved with the help of Nollert's idea~\cite{Nollert:1993zz,Zhidenko:2006rs}. It consists in using an estimate for the ratio of the series coefficients at large $n$ to truncate the continued fraction equation. From the explicit terms of the recurrence relation \eqref{eq:5-terms}, we find that, up to order $n^{-1}$, the convergent series satifies
\begin{align} 
 \frac{a_{n+1}}{a_n}  & = 1 - \sqrt{\frac{-2q(r_0-r_1)}{n}} \nonumber \\ & -  \left[\frac{3}{4} + q(r_0-r_1+M) - \frac{M\omega^2}{q} \right]\frac{1}{n} . \label{nolimp1}
\end{align}

An alternative \emph{ansatz} to \eqref{eq:radial_solution_KZ}, which also yields a five-term recurrence relation, is given by
\begin{align}
	R\left( r\right)&=e^{qr}\left( r-r_{0}\right) ^{-i \sigma_{0}}(r-r_{1})^{-i \sigma_{1}}(r-r_{2})^{-i \sigma_{2}}  \nonumber \\  & \times   r^{-1+\chi + 2 iM \omega} \sum ^{\infty }_{n=0}b_{n}\left( \frac{r-r_{0}}{r}\right) ^{n}, \label{eq:radial_solution_KZ2}
\end{align}      
where $b_n \in \mathbb{C}$. The important difference in  comparison to \eqref{eq:radial_solution_KZ} is that the M\"obius transformation associated with \eqref{eq:radial_solution_KZ2}, given by $u=(r-r_0)/r$, now takes the singular points $r=0$, $r=r_1$ and $r=r_2$, respectively into
\begin{equation}
u=\infty, \quad u=u_1=\frac{r_1-r_0}{r_1}, \quad u=u_2 = \frac{r_2-r_0}{r_2}.
\end{equation}   
Hence, the continued fraction method based on $\eqref{eq:radial_solution_KZ2}$ is applicable if $|u_1|>1$ and $|u_2|>1$. The region of the parameter space where these inequalities are satisfied is indicated by the white dashed curve in Fig.~\ref{fig:horizons}. We highlight that the use of both \emph{ansatzes} \eqref{eq:radial_solution_KZ} and \eqref{eq:radial_solution_KZ2} increase the applicability of the continued fraction method.

For completeness, we present below the large-$n$ limit of the ratio $b_{n+1}/b_n$ obtained from the recurrence relation associated with \eqref{eq:radial_solution_KZ2}. Omitting corrections of order higher than $n^{-1}$, we obtain
\begin{align}
 \frac{b_{n+1}}{b_n}  & = 1 - \sqrt{\frac{-2qr_0}{n}} \nonumber \\ & -  \left[\frac{3}{4} + q(r_0+M) - \frac{M\omega^2}{q} \right]\frac{1}{n} . \label{nolimp2}
\end{align}
 As before, this expression is used to implement Nollert's improvement of the continued fraction method.
           

\section{Numerical Results}
We use the continued method fraction described in the previous section to determine the QNMs and the QBSs of a massive scalar field in the deformed Kerr black hole \eqref{eq:deformed_Kerr_line_element}. For a given set of parameters $M$, $a$, $\eta$, $\mu$, $\ell$, and $m$, the continued fraction equation \eqref{cfequ} [or its alternative version obtained from \eqref{eq:radial_solution_KZ2}] depends only on the eigenfrequency $\omega$. To determine $\omega$ we truncate the continued fraction at $N$ terms and solve the equation using a root finding algorithm.
We use Nollert's improvement \eqref{nolimp1} [or its alternative version \eqref{nolimp2}].

In each step we start with $N=100$ and repeat the calculation, increasing $N$ by $100$, until the relative difference between the last and the second to last values for the root reach the specified threshold $\epsilon = 10^{-7}$. More precisely, if $\omega_{(N)}$ denotes the eigenfrequency found with an $N$-term continued fraction, we stop the iteration and set $\omega=\omega_{(N)}$ when
\begin{equation}
\left| \frac{\omega_{(N)} - \omega_{(N-100)}}{\omega_{(N)}}\right| < \epsilon. 
\end{equation} 
Previous calculations of eigenfrequencies in the Kerr background~\cite{Leaver1985,Dolan2007,Zhidenko2006}  
are used to validate our implementation of the continued fraction method. 

To scan the space $\left(a/M,\eta/M\right)$ of black hole parameters where the continued fraction method is applicable (see Fig.~\ref{fig:horizons}), we fix $\mu M$, $\ell$, and $m$ and vary $a/M$ and $\eta/M$ incrementally. The starting point of the calculations is the point $(0.001,0)$ of the parameter space, which corresponds to a slowly rotating Kerr black hole. Using the Schwarzschild fundamental mode as the initial guess for the root finding algorithm, we move up in the parameter space in steps of $0.01$ until the point $(0.001,3)$ is reached. At each step, the root found in the previous step is used as the initial guess for the eigenfrequency. From $(0.001,3)$, we move right in steps of $0.01$ until the point $(1.991,3)$ is reached. From each point in the $\eta/M = 3$ line, we move down in the parameter space until the whole region determined by the white and red curves of Fig.~\ref{fig:horizons} has been covered.

\subsection{Quasinormal modes}
\begin{figure}[!htbp]
\centering
  \includegraphics[width = 0.9 \linewidth]{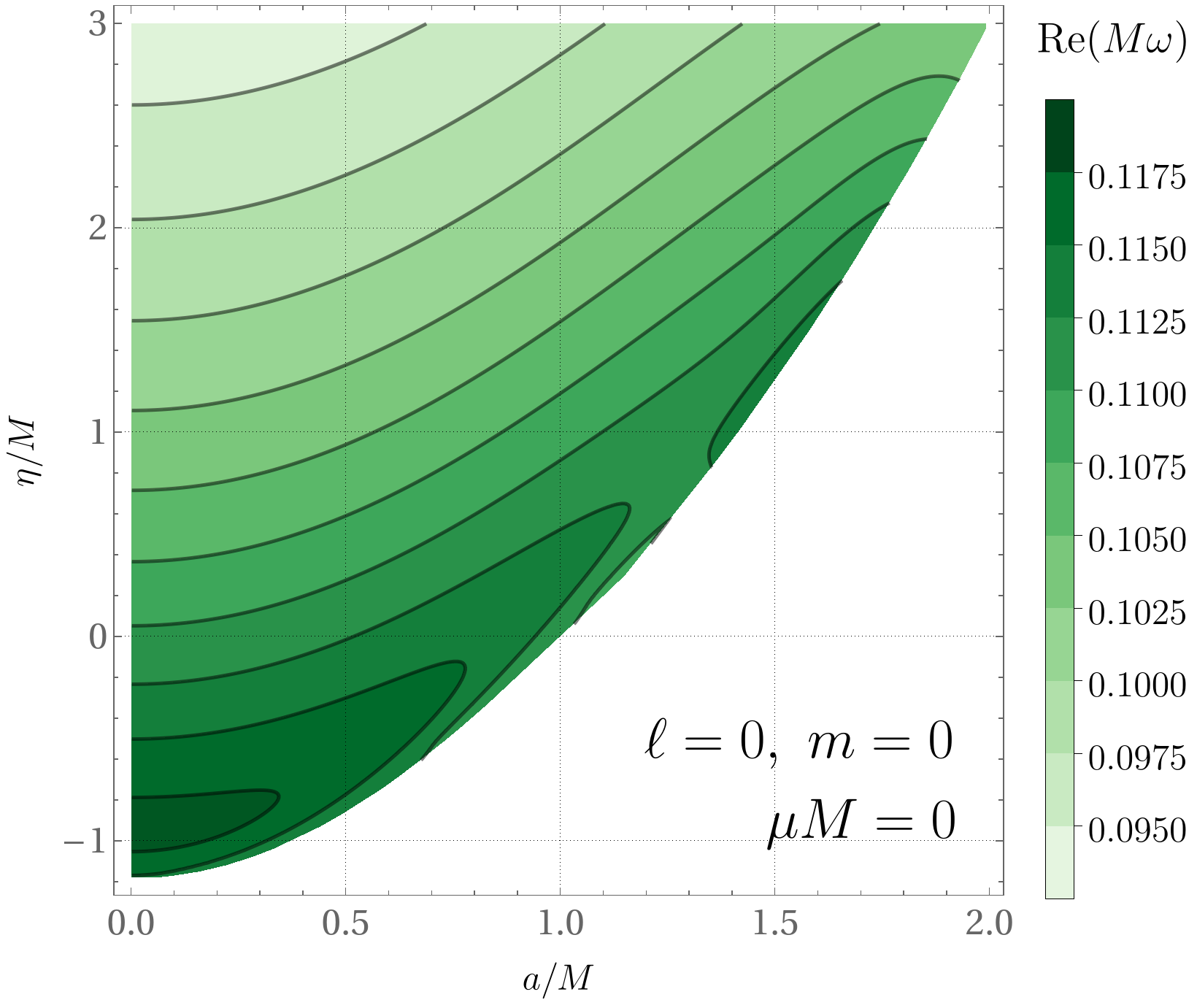} 
  \includegraphics[width = 0.9 \linewidth]{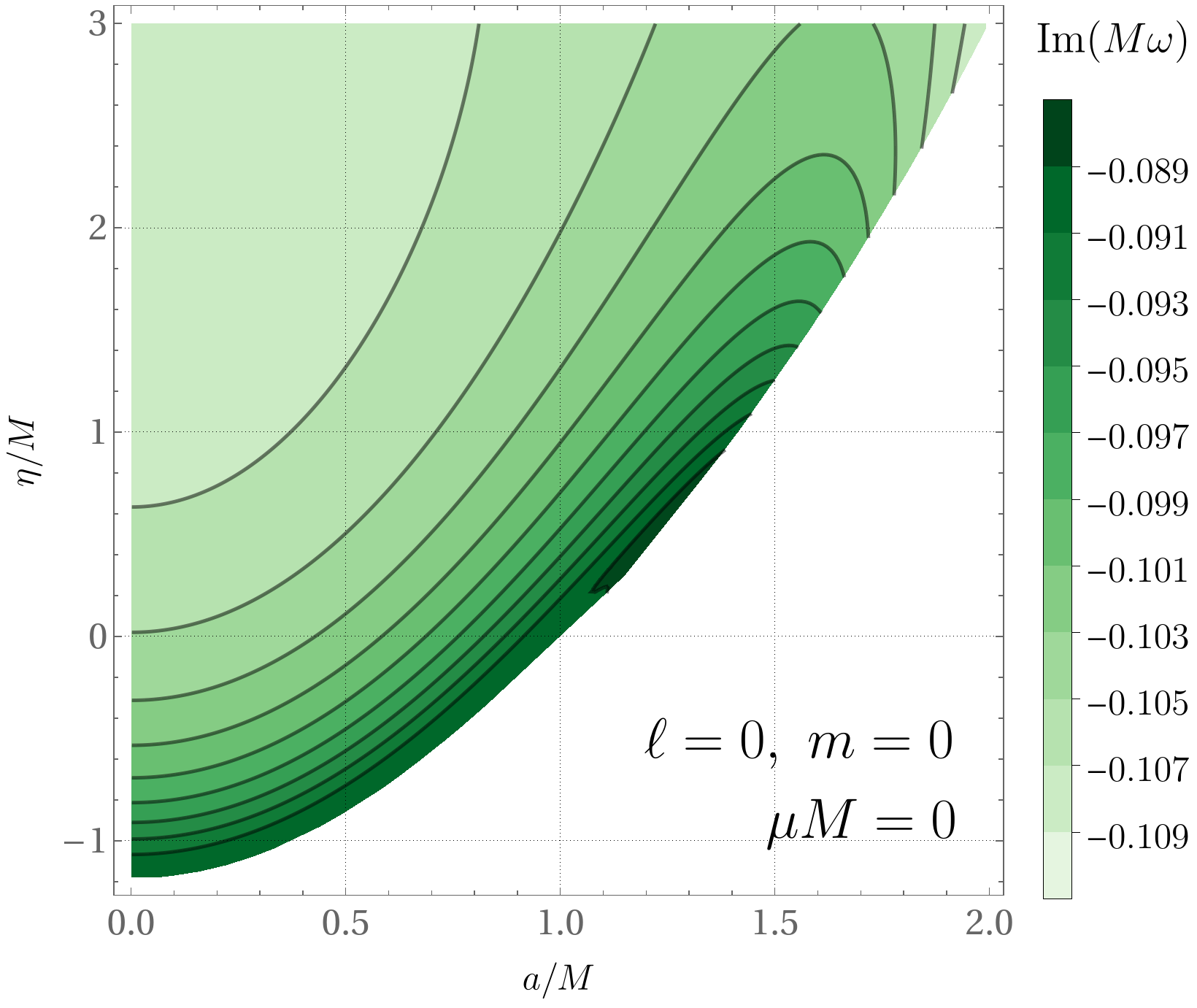}
  \caption{Real (top panel) and imaginary (bottom panel) parts of the QNM fundamental frequency of the $\ell=m=0$  massless scalar perturbations of the deformed Kerr black hole as a function of the spin $a/M$ and the deformation $\eta/M$. }
  \label{fig:qnmreiml0m0} 
\end{figure}

We first investigate the behaviour of a massless scalar perturbation around a deformed Kerr metric.
To understand the dependence of the fundamental QNM on the spin $a/M$ and on the deformation $\eta/M$, we have scanned the parameter space using the routine sketched above. The results for the $\ell=m=0$ modes are shown in Fig.~\ref{fig:qnmreiml0m0}, where we plot the real (top panel) and imaginary (bottom panel) parts of the QNM frequency. The contour lines indicate that the slowest oscillation rates are associated with the top left corner of the parameter space (i.e.~small $a/M$ and large $\eta/M$). Such modes also have the shortest decay times in the parameter space under analysis. The fastest oscillation rates, on the other hand, occur for $\eta/M \sim -1$ and small values of spin ($a/M \lesssim 0.3 $). The least stable modes occupy the region around the triple point $(2/\sqrt{3},8/27)$.

\begin{figure}[!htbp]
\centering
  \includegraphics[width = 0.95 \linewidth]{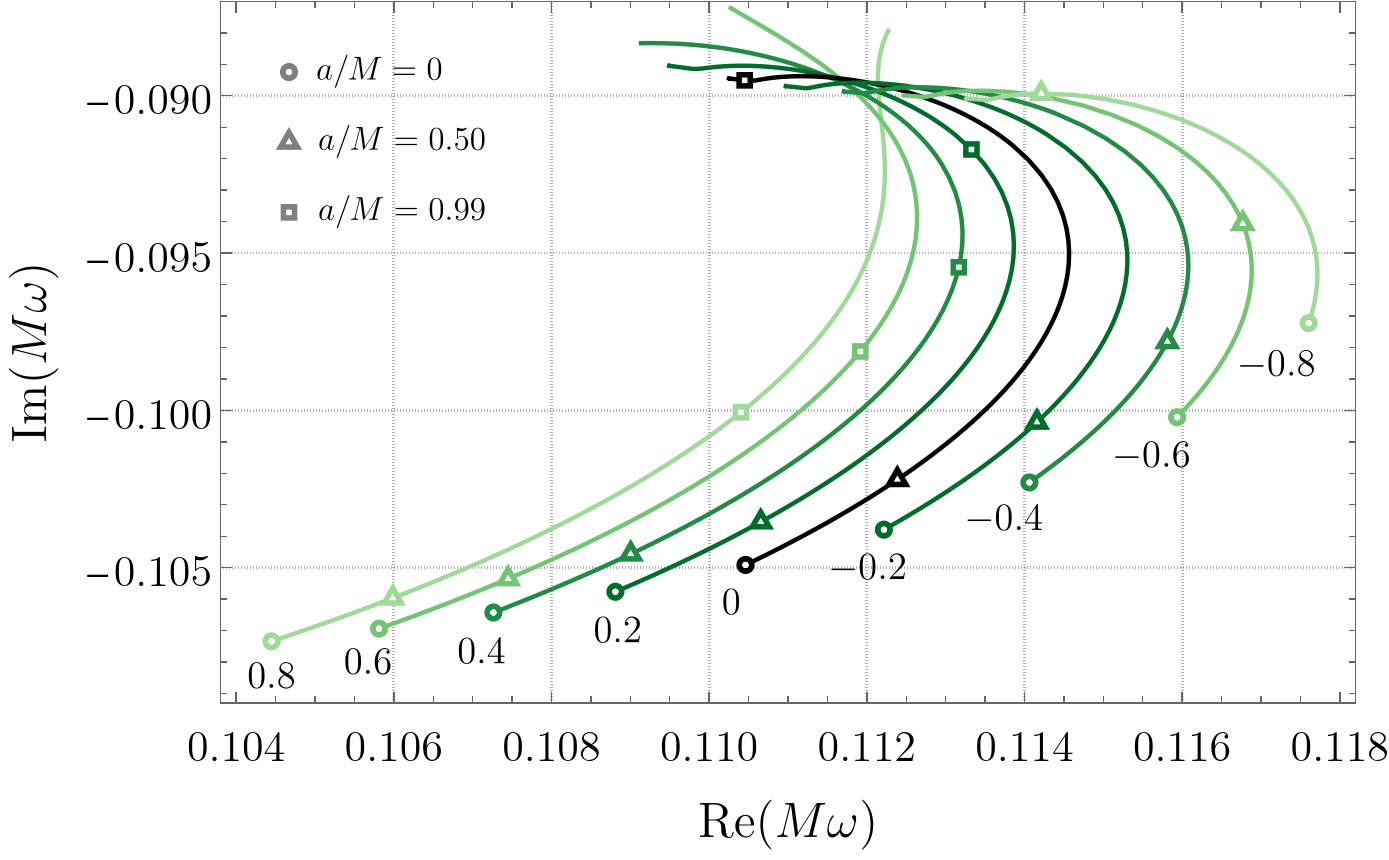}
  \caption{Parametric plot of the QNM fundamental frequency of the $\ell=m=0$  massless scalar perturbations of the deformed Kerr black hole for selected values (shown below each curve) of $\eta/M$. In each curve, the parameter $a/M$ varies from zero until the corresponding maximum value given in Fig.~\ref{fig:qnmreiml0m0}. The black line corresponds to Kerr black holes.}
  \label{fig:qnml0m0} 
\end{figure}

The comparison between the $\ell=0$ massless QNMs of the Kerr and the deformed Kerr black holes is better visualized in Fig.~\ref{fig:qnml0m0}. It shows a parametric plot of the real and imaginary parts of the frequency, as a function of the spin $a/M$, for selected values of the deformation parameter $\eta/M$.     The chosen values, $\eta/M=-0.8,-0.6,\dots,0.6,0.8$, are specified below each curve. Black curves correspond to Kerr black holes. Green curves, on the other hand, correspond to deformed Kerr black holes (darker tones represent smaller deformations, while lighter tones represent larger deformations). In particular, for a fixed small spin $a/M$, small positive deformations from Kerr make the corresponding QNMs oscillate slower and decay faster. In contrast, for a rapidly rotating Kerr black hole with $a \sim 1$, small positive deformations $\eta$ make the corresponding QNMs oscillate and decay faster.

\begin{figure*}[!htbp]
\centering
  \includegraphics[width = 0.9 \linewidth]{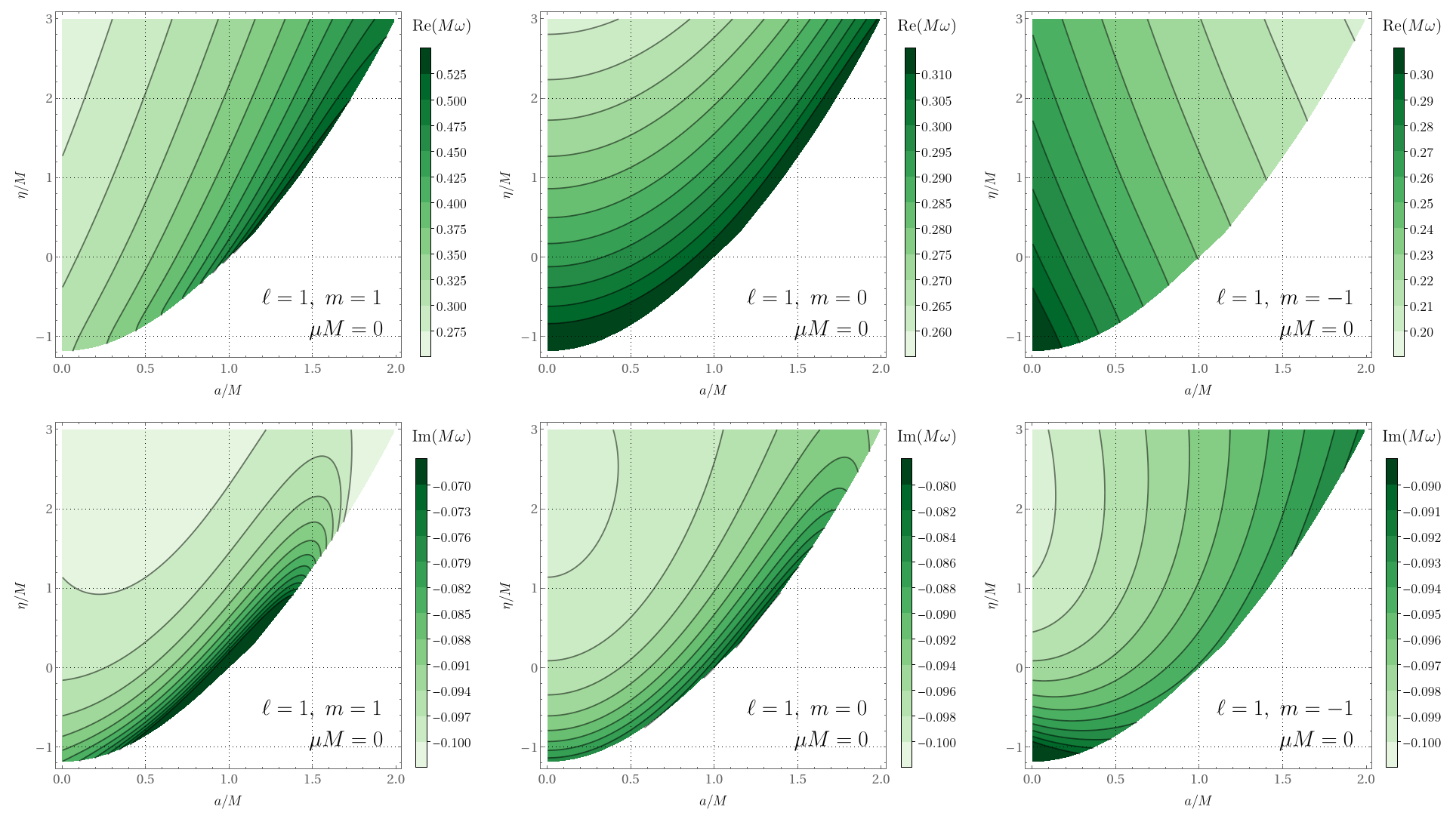}
  \caption{Real (top row) and imaginary (bottow row) parts of the fundamental QNM frequency of the $\ell=1$ massless scalar perturbations of the deformed Kerr black hole as a function of the spin $a/M$ and the deformation $\eta/M$. The left panels correspond to $m=1$, the middle ones to $m=0$, and the right ones to $m=-1$. 
 }
  \label{fig:qnmreiml1} 
\end{figure*}

\begin{figure}[!htbp]
\centering
  \includegraphics[width = 0.9 \linewidth]{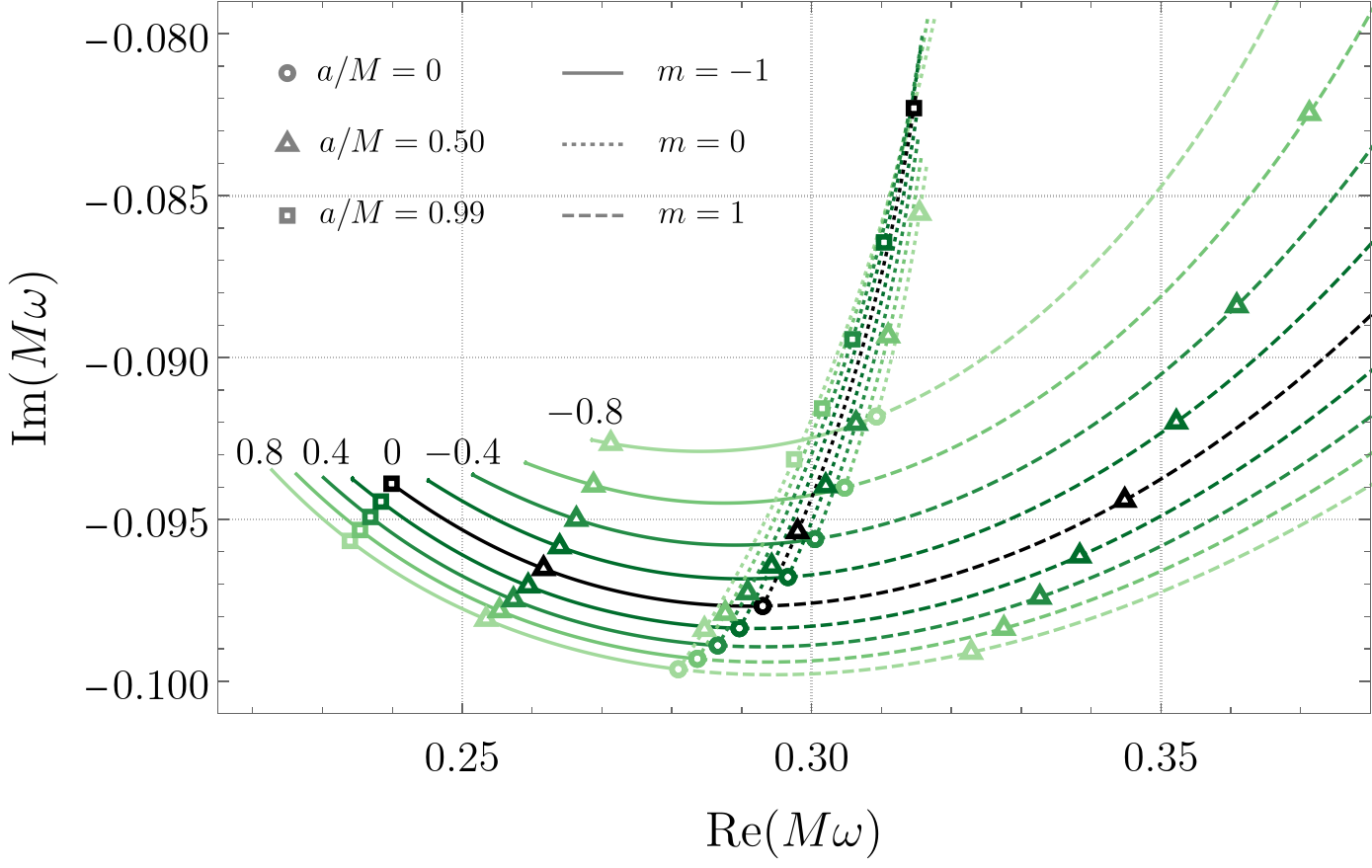}
  \includegraphics[width = 0.9 \linewidth]{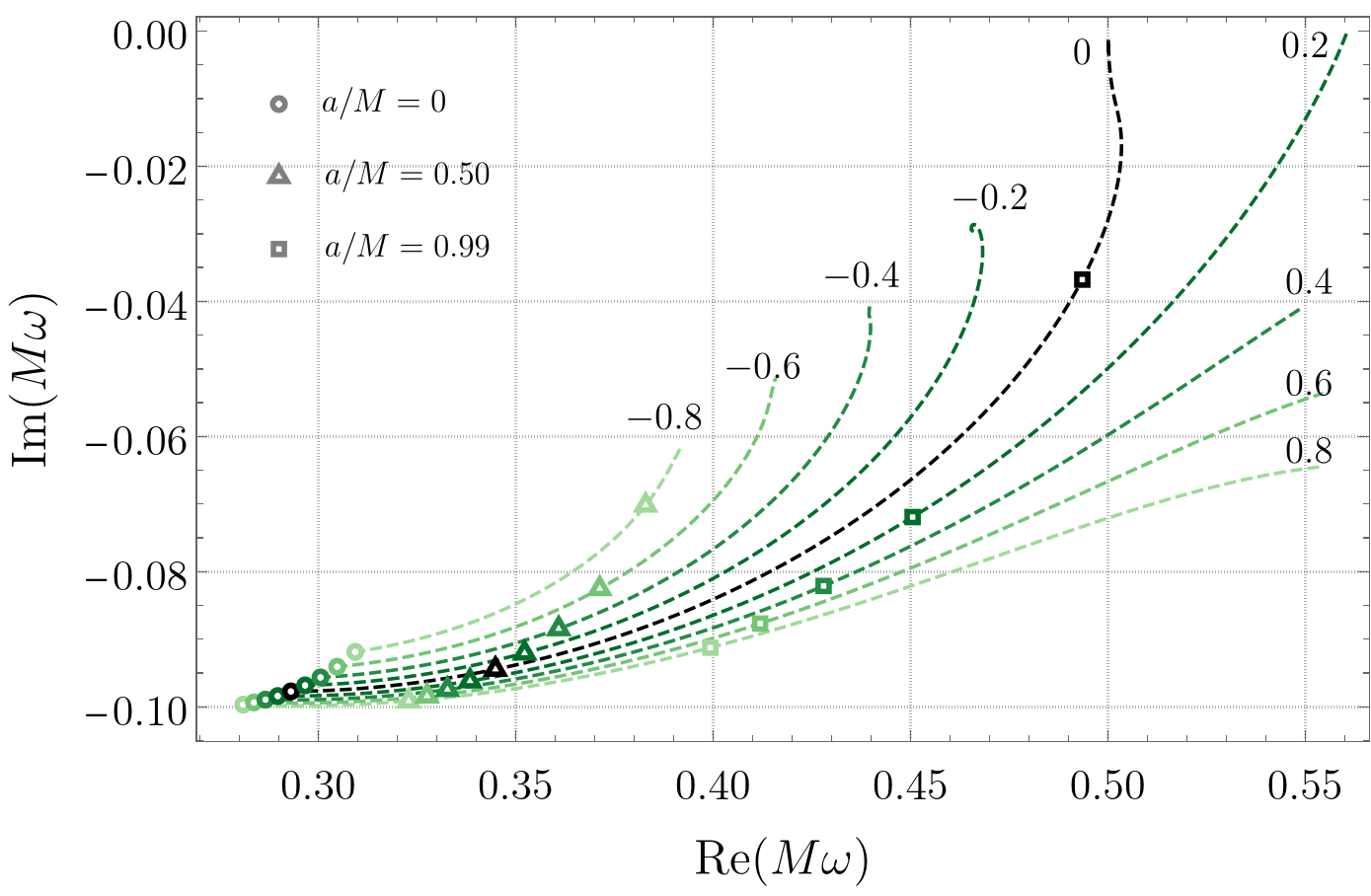}
  \caption{Parametric plots of the $\ell=1$ QNMs for selected values (shown besides each curve) of $\eta/M$. The top panel focuses on the $m=-1$ and $m=0$ modes, while the bottom panel shows the full range of the $m=1$ curves. 
}
  \label{fig:qnml1} 
\end{figure}

Our analysis of the QNMs of a massless scalar field is complemented by the results for the $\ell=1$ modes shown in Fig.~\ref{fig:qnmreiml1}. We exhibit the real and imaginary parts of the QNM frequency for $m=1$ (left panels), $m=0$ (middle panels), and $m=-1$ (right panels). With respect to the $\ell=0$ modes, the $\ell=1$ modes typically have higher oscillation frequencies in the entire parameter space under investigation. For both $m=0$ and $m=1$, we observe that the real part of the frequency increases when $\eta/M$ decreases and $a/M$ increases. For $m=-1$, on the other hand, $\mathrm{Re}(M\omega)$ increases when both $\eta/M$ and $a/M$ decrease. Regarding the decay rates, we observe that the least stable modes are concentrated around the triple point for $m=1$ and $m=0$ (as in the case of the $\ell = 0$ modes), while for $m=-1$ they occupy the bottom left and the top right corners of the parameter space. 

In Fig.~\ref{fig:qnml1} we examine the dependence of the $\ell=1$ QNMs on the spin parameter $a/M$ for the same selection of deformation parameters $\eta/M$ used in Fig.~\ref{fig:qnml0m0}. As before, black curves correspond to Kerr black holes and green curves correspond to deformed Kerr black holes (darker tones represent smaller deformations, while lighter tones represent larger deformations). At fixed spin, positive $\eta$ deformations from Kerr typically make the corresponding QNMs oscillate slower and decay faster, while negative ones make them oscillate faster and decay slower.

\begin{figure*}[!htbp]
\centering
  \includegraphics[width = 0.95 \linewidth]{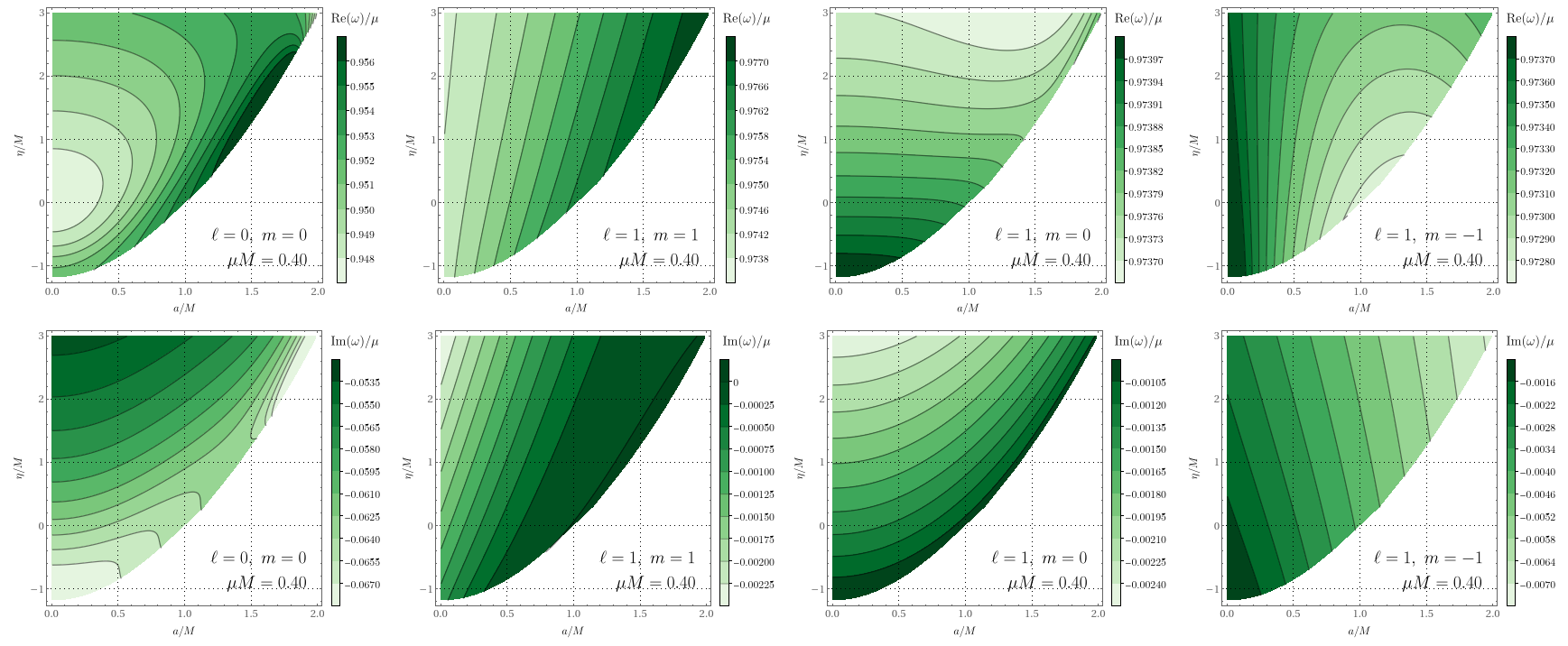}
  \caption{Real (top row) and imaginary (bottow row) parts of the QBS fundamental frequency of the $\ell=0$ and $\ell=1$ massive ($\mu M = 0.4$) scalar perturbations of the deformed Kerr black hole. In each plot the horizontal and vertical axes are, respectively, the spin $a/M$ and the deformation $\eta/M$ of the black hole.}
  \label{fig:qbs40} 
\end{figure*}

In particular, if $m=-1$ or $m=0$, we note that $\mathrm{Im}(M\omega)$ tends to a finite negative value as the maximum spin is approached. The same behavior is observed in the case $m=1$ for most deformation parameters. However, when $0 \le \eta/M \le 8/27$, our results suggest that the imaginary frequency of the co-rotating modes approaches zero in the extremal black hole limit. The modes thus become marginally stable in this limit. Fig.~\ref{fig:qnml1} exhibits this explicitly for $\eta/M=0$ and $\eta/M=0.2$. We note that the corresponding oscillation frequencies tend to $m\Omega_0$ as extremality is approached in these cases. Such modes, known as zero-damped modes, have been investigated by several authors in the past for nearly extreme and extreme Kerr black holes~\cite{1980ApJ...239..292D,Glampedakis:2001js,Yang:2012pj,Richartz:2015saa,Richartz:2017qep}. Our results demonstrate the existence of zero-damped modes for the deformed Kerr black hole \eqref{eq:deformed_Kerr_line_element}.

\subsection{Quasibound states and superradiant instabilities}
The previous analysis on QNMs assumed a massless scalar field. We now consider a massive field and analyze the existence of QBSs. Following the same routine used for QNMs, we scan the space $\left(a/M,\eta/M\right)$ in order to investigate the dependence of the fundamental QBS on the black hole parameters. To illustrate the general properties of the QBSs of the deformed Kerr black hole, we choose, as an example, $\mu M =0.4$. The results for $\ell=0$ and $\ell=1$ are shown in Fig.~\ref{fig:qbs40}. The plots in the top row of the figure show that the mass of the scalar field is slightly larger than the real part of the frequency, which is a typical feature of QBSs. Comparing the top row with the bottow row, we also deduce that the real part of the frequency is typically much larger than its imaginary part in the region of interest. Comparison with Figs.~\ref{fig:qnmreiml0m0} and \ref{fig:qnmreiml1} shows that the decay times of the QBSs of a massive scalar field with $\mu M = 0.4$ are longer than the corresponding decay times of the QNMs of a massless scalar field.

We further observe that the modes shown in Fig.~\ref{fig:qbs40} are typically characterized by $\mathrm{Im}(M\omega)<0$ and, thus, are stable. However, corotating modes may by associated with superradiant instabilities and, in such cases, will be characterized by $\mathrm{Im}(M\omega)>0$. In Fig.~\ref{fig:qbs40}, superradiant instabilities can be identified in the darkest part of the $\ell=m=1$ panel of the bottow row. For this particular mass  $\mu M=0.4$, we see that instabilities are possible if $\eta/M \gtrsim -0.4$. We remark that the contour line $\mathrm{Im}(M\omega)=0$, which separates the time-decaying modes from the superradiant instabilities, correspond to bound states. Such bound states are precisely the scalar clouds defined by Eq.~\eqref{scalarcloudsdef}. 

\begin{figure}[!htbp]
\centering
  \includegraphics[width = 0.95 \linewidth]{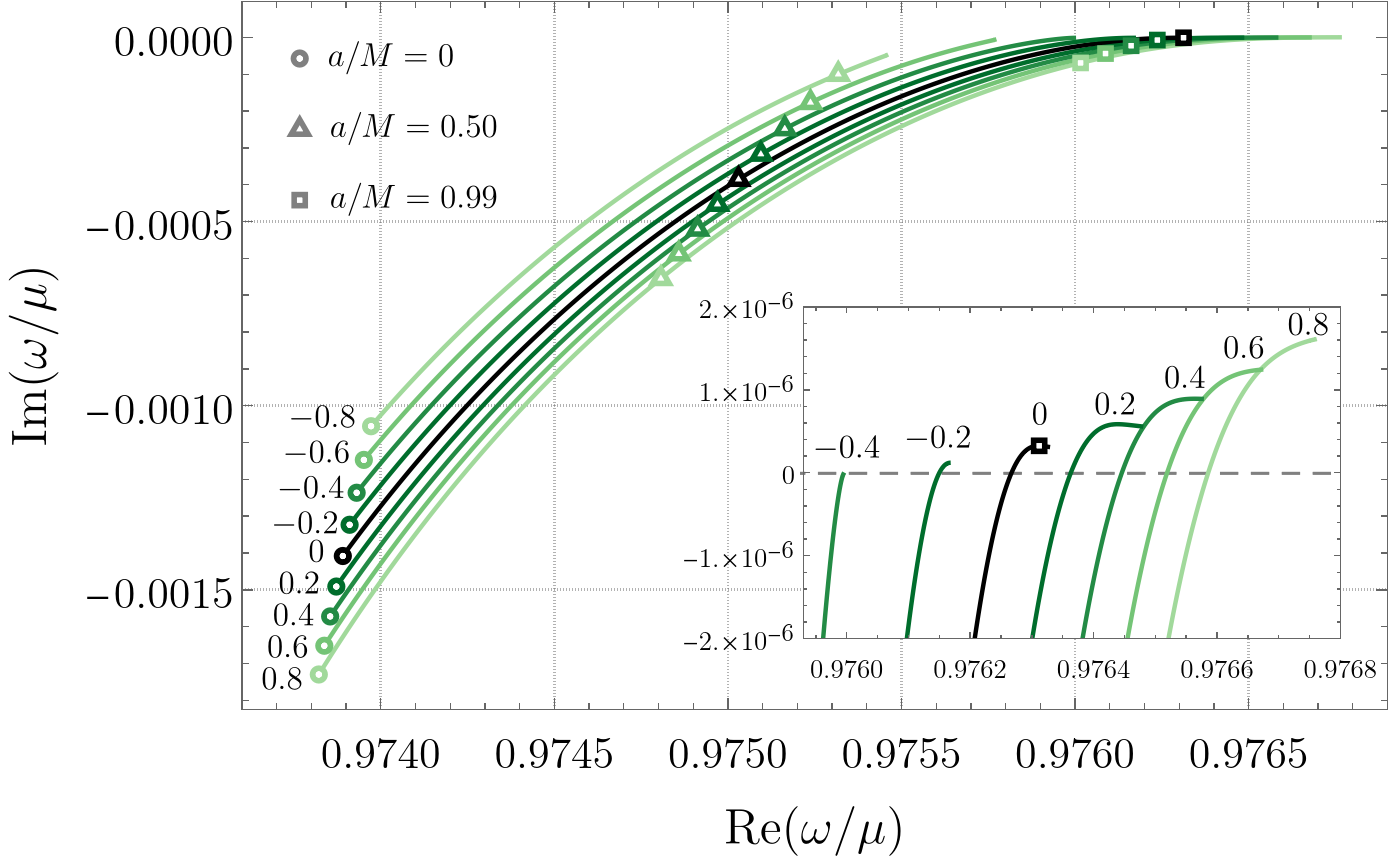}
  \caption{QBSs of the $\ell=m=1$  massive ($\mu M = 0.4$) scalar perturbations  of the deformed Kerr black hole for selected values (shown next to each curve) of the deformation parameter $\eta/M$. The inset highlights the regime of superradiant instabilities. The points at which the curves intersect the dashed horizontal line correspond to scalar clouds. 
 In each curve, the parameter $a/M$ varies from zero until the corresponding maximum value given in Fig.~\ref{fig:qbs40}. The black curve corresponds to Kerr black holes.}
  \label{fig:qbs40reim} 
\end{figure}

In Fig.~\ref{fig:qbs40reim} we plot the $\ell=m=1$ QBSs for a selection of $\eta/M$ values around zero (Kerr black hole). The chosen values, $\eta/M=-0.8,-0.6,\dots,0.6,0.8$, are specified next to each curve. Black curves correspond to Kerr black holes. Green curves correspond to deformed Kerr black holes (darker tones represent smaller deformations, while lighter tones represent larger deformations). In each curve, the spin varies from zero until the corresponding maximum given in Fig.~\ref{fig:qbs40}. We observe that, similarly to the analysis for QNMs, when $a/M$ is fixed, positive $\eta$ deformations from Kerr decrease both the oscillation rate and the imaginary part of the frequency.

 The inset of Fig.~\ref{fig:qbs40reim} exhibits in detail the superradiantly unstable modes, for which $\mathrm{Im}(M\omega)>0$, and the scalar clouds, for which $\mathrm{Im}(M\omega)=0$. We see that, as the deformation parameter increases, the peak of the curves, corresponding to the maximum instability, also increases. The spin parameter at which such maximum instability occurs also increases with the deformation parameter. Regarding the scalar clouds for fixed $\mu M = 0.4$, the inset of Fig.~\ref{fig:qbs40reim} shows that the associated oscillation frequencies $\omega_{\mathrm{sc}}$ (and the spin parameter at which they occur) also increase when the deformation parameter increases.

\begin{figure}[!htbp]
\centering
  \includegraphics[width = 0.95 \linewidth]{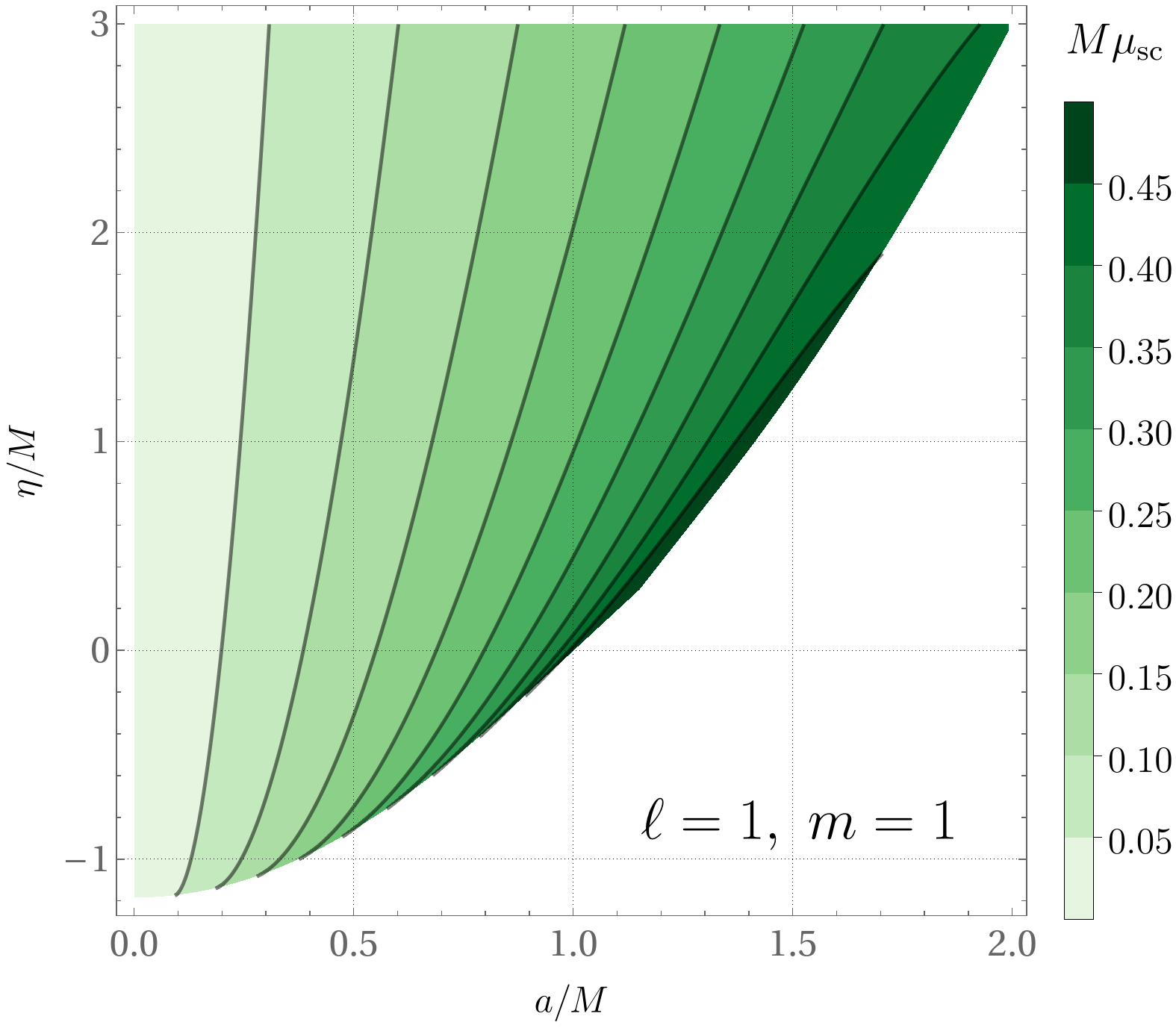}
  \includegraphics[width = 0.95 \linewidth]{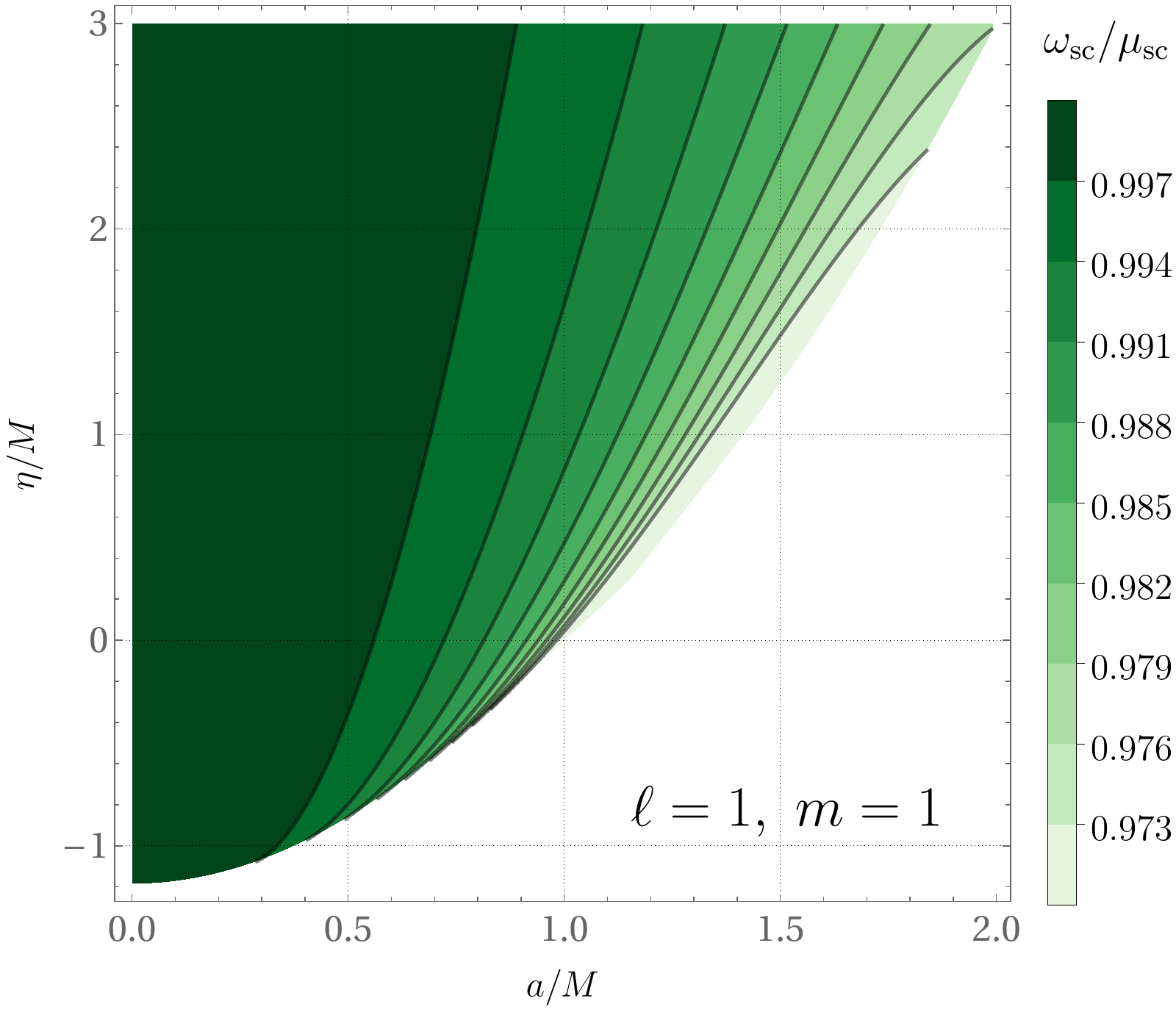} 
  \caption{Mass $M\mu_{\mathrm{sc}}$ (top panel) and the ratio between oscillation frequency and mass $\omega_{\mathrm{sc}}/\mu_{\mathrm{sc}}$ (bottom panel)  of the fundamental $\ell=m=1$ scalar cloud of the deformed Kerr black hole as a function of the spin $a/M$ and the deformation parameter $\eta/M$.}
  \label{fig:mu_sc} 
\end{figure}

\subsection{Scalar clouds}
In the previous section we explained how the values of spin and deformation for which the scalar clouds occur can be determined by looking at the transition from stability to instability at fixed $\mu M$. We now investigate the properties of scalar clouds as the mass parameter $\mu M$ varies. However, instead of looking for transitions, we adopt a more straightforward approach that requires going through the parameter space $\left(a/M,\eta/M\right)$ only one time (as opposed to once for each choice of $\mu M$).

\begin{figure}[!htbp]
\centering
    \includegraphics[width = 0.95 \linewidth]{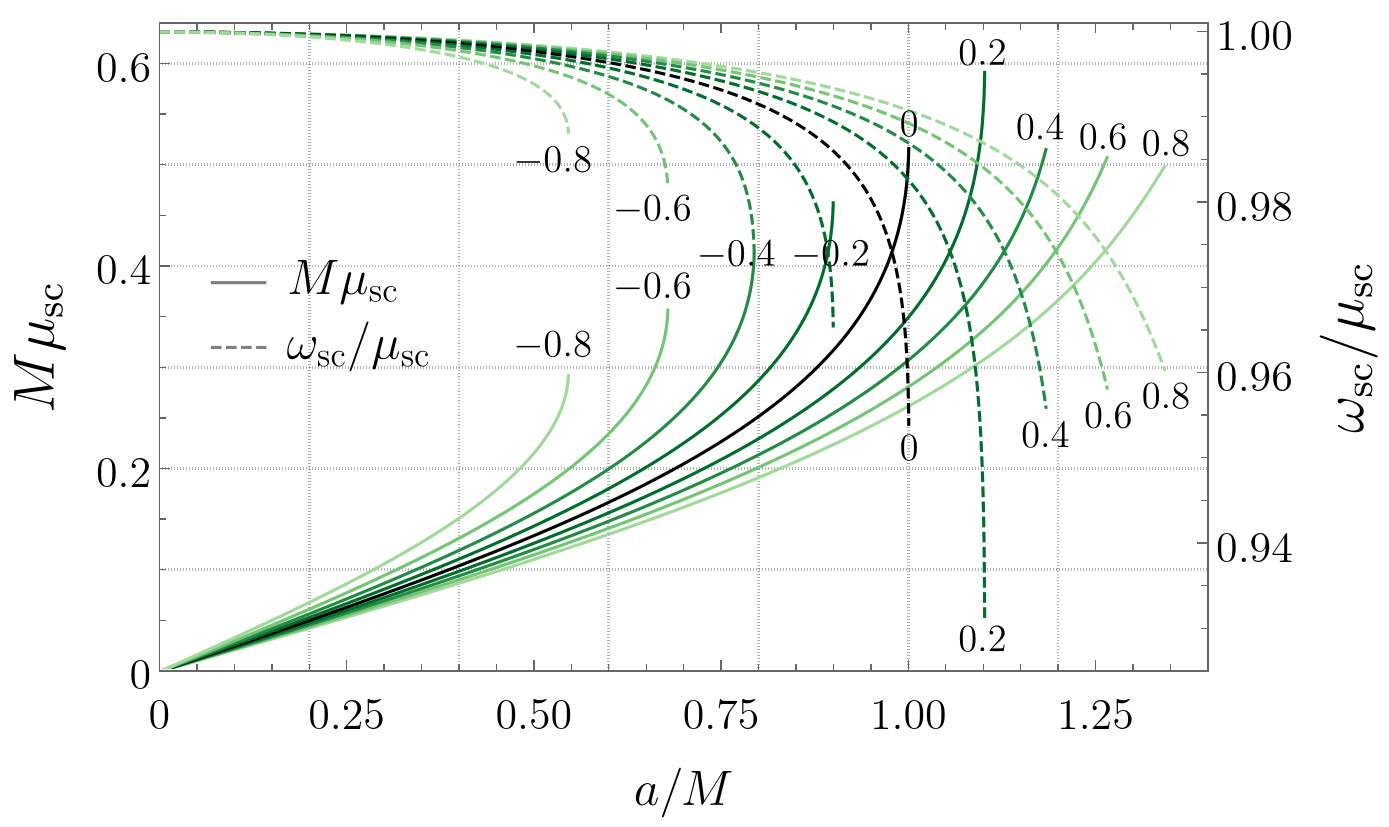}
  \caption{Mass $M\mu_{\mathrm{sc}}$ (solid curves) and the ratio between oscillation frequency and mass $\omega_{\mathrm{sc}}/\mu_{\mathrm{sc}}$ (dashed curves) of the fundamental $\ell=m=1$ scalar cloud of the deformed Kerr black hole as a function of the spin $a/M$ for selected values (shown next to each curve) of the deformation parameter $\eta/M$. In each curve, the parameter $a/M$ varies from zero until the corresponding maximum value given in Fig.~\ref{fig:mu_sc}. The black curves correspond to Kerr black holes.}
  \label{fig:mu_sc2} 
\end{figure}

 The procedure we use to determine scalar clouds fixes $\omega = m \Omega_0$ by assumption and takes $\mu$ to be the unknown parameter. In other words, for each set of parameters ($M$, $a$, $\eta$, $\ell$, and $m$), after $\omega = m \Omega_0$ is determined through \eqref{Omega0}, the continued fraction equation depends only on $\mu$. Let $\mu_{(N)}$ denote the solution found when the continued fraction is truncated at $N$ terms. As before, we start with $N=100$ and repeat the calculation, increasing $N$ by $100$ until 
\begin{equation}
\left| \frac{\mu_{(N)} - \mu_{(N-100)}}{\mu_{(N)}}\right| < \epsilon, 
\end{equation}
at which point we set $\mu=\mu_{(N)}$. Previous calculations of the mass of the scalar field  in the Kerr background~\cite{Hod:2012px,Herdeiro:2014goa,Benone:2014ssa} are used to validate our implementation of the continued fraction method. 

\begin{figure*}[!htbp]
\centering
  \includegraphics[width = 0.95 \linewidth]{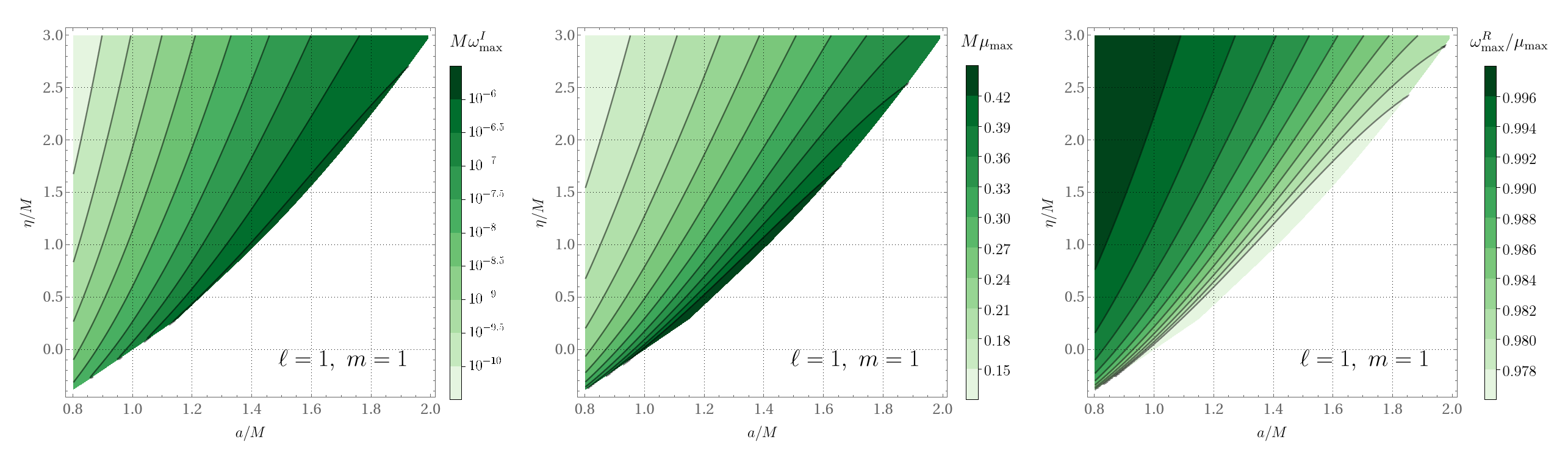}
  \caption{The maximum superradiant instability $M\omega^{I}_{\mathrm{max}}$ (left panel), the corresponding mass $M\mu_{\mathrm{max}}$ (middle panel), and the associated ratio $\omega^{R}_{\mathrm{max}}/\mu_{\mathrm{max}}$ between the oscillation frequency and the mass (right panel). In each plot the horizontal and vertical axes are, respectively, the spin parameter $a/M$ and the deformation parameter $\eta/M$ of the black hole. The parameters associated with the scalar field are $\ell=m=1$.}
  \label{fig:max_ins} 
\end{figure*}

To scan the space $\left(a/M,\eta/M\right)$ of black hole parameters where the continued fraction method is applicable (see Fig.~\ref{fig:horizons}), we fix $\ell$ and $m$, and vary $a/M$ and $\eta/M$ incrementally. The starting point of the calculations is the point $(0.991,0)$ of the parameter space, which corresponds to a rapidly rotating Kerr black hole. 
We then move up in the parameter space in steps of $0.01$ until the point $(0.991,3)$ is reached. At each step, we compute $\omega=m\Omega_0$ and use the root found in the previous step as the initial guess for the eigenfrequency. From $(0.991,3)$, we move right and left in steps of $0.01$ until the points $(0.001,3)$ and $(1.991,3)$ are reached. From each point in the $\eta/M = 3$ line, we move down in the parameter space until the whole region determined by the white and red lines of Fig.~\ref{fig:horizons} has been covered.

The results for the fundamental $\ell=m=1$ scalar cloud of the deformed Kerr black hole are displayed in Fig.~\ref{fig:mu_sc}. The top and bottom panels show, respectively, the mass $M\mu_{\mathrm{sc}}$ and the ratio  $\omega_{\mathrm{sc}}/\mu_{\mathrm{sc}}$ as a function of the black hole parameters. We note that, when the deformation parameter is fixed, the mass of the scalar cloud typically increases (and the ratio $\omega_{\mathrm{sc}}/\mu_{\mathrm{sc}}$ decreases) when the spin parameter increases.
In contrast, for a given spin, the mass of the scalar cloud typically decreases (and the ratio $\omega_{\mathrm{sc}}/\mu_{\mathrm{sc}}$ increases) when the deformation parameter increases. The comparison with Kerr is more evident in Fig~\ref{fig:mu_sc2}, where we exhibit the mass $M\mu_{\mathrm{sc}}$ (solid curves) and the ratio $\omega_{\mathrm{sc}}/\mu_{\mathrm{sc}}$ (dashed curves) as a function of the spin $a/M$ for selected values of the deformation parameter $\eta/M$.

Figs.~\ref{fig:mu_sc} and \ref{fig:mu_sc2} also suggest that there is a maximum mass parameter $\mu M$ above which scalar clouds (and, therefore, superradiant instabilities) are turned off. This threshold has been previously investigated for Kerr black holes taking into account all azimuthal and orbital numbers~\cite{beyer1,beyer2,hod1,hod2}. Our numerical results show that, for $\ell = m =1$, the maximum mass that allows scalar clouds is associated with the triple point $(2/\sqrt{3},8/27)$.

\subsection{Maximum instability}
We also investigate the maximum instability of the $\ell=m=1$ perturbations as the black hole parameters vary. In order to do so, we use the results discussed in the previous section (and displayed in Fig.~\ref{fig:mu_sc}) as initial points in our analysis. More precisely, for each pair $\left(a/M,\eta/M \right)$ we set the mass and the frequency of the scalar field to be the corresponding ones for the scalar cloud, i.e.~$\mu=\mu_{\mathrm{sc}}$ and $\omega=\omega_{\mathrm{sc}}=m\Omega_0$. We then decrease the mass $\mu$ incrementally, employing the continued fraction method to determine the eigenfrequency $\omega$ at each step. The initial guess for the root finding algorithm is the eigenfrequency found in the previous step. For each pair of black hole parameters, we stop the iteration as soon as a local maximum of $\mathrm{Im}(M\omega)$ (as a function of $\mu$) is identified. The peak value $\mathrm{Im}(\omega)$ is denoted $\omega^{I}_{\mathrm{max}}$ while the associated oscillation frequency $\mathrm{Re}(\omega)$ and the associated mass $\mu$ are denoted, respectively, by $\omega^{R}_{\mathrm{max}}$ and $\mu_{\mathrm{max}}$. Since it becomes more difficult to identify the peaks as the spin decreases, we focus here on the $a/M > 0.8$ region of the parameter space.

The results are plotted in Fig.~\ref{fig:max_ins}: from left to right, the panels exhibit the maximum instability $M\omega^{I}_{\mathrm{max}}$, the corresponding mass $M\mu_{\mathrm{max}}$, and the associated ratio $\omega^{R}_{\mathrm{max}}/\mu_{\mathrm{max}}$ as a function of the black hole parameters. We verify that, typically, both the maximum instability $M\omega^{I}_{\mathrm{max}}$ and the associated scalar field mass $M\mu_{\mathrm{max}}$ increase when $a/M$ increases (for fixed $\eta/M$).  
Similarly, when the spin is fixed, both $M\omega^{I}_{\mathrm{max}}$ and $M\mu_{\mathrm{max}}$ typically increase when the deformation parameter decreases. On the other hand, the ratio $\omega^{R}_{\mathrm{max}}/\mu_{\mathrm{max}}$ increases towards the top left corner of the parameter space, i.e.~when the spin parameter decreases and the deformation parameter increases.  

\begin{figure}[!htbp]
\centering
  \includegraphics[width = 0.95 \linewidth]{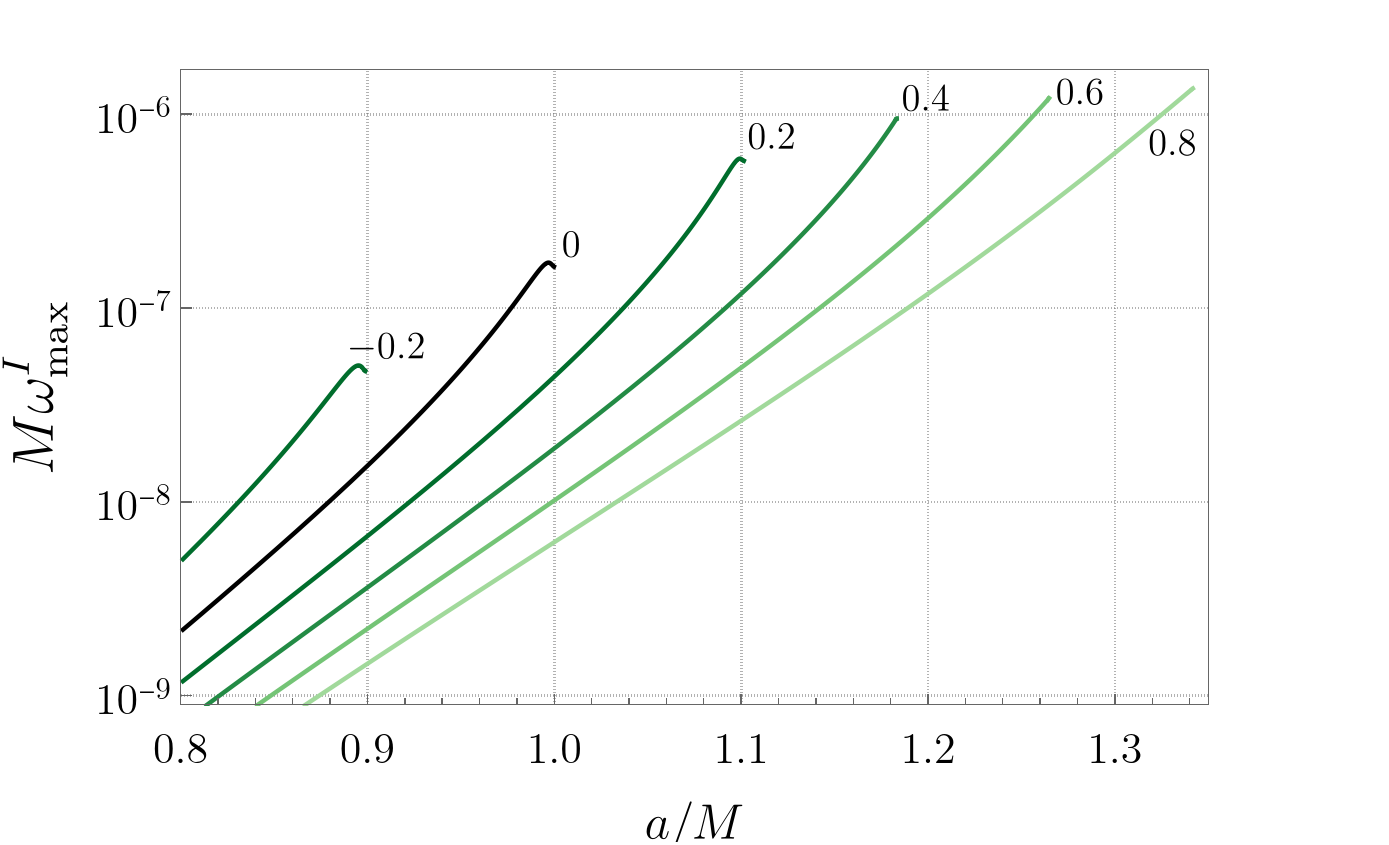}
  \includegraphics[width = 0.95 \linewidth]{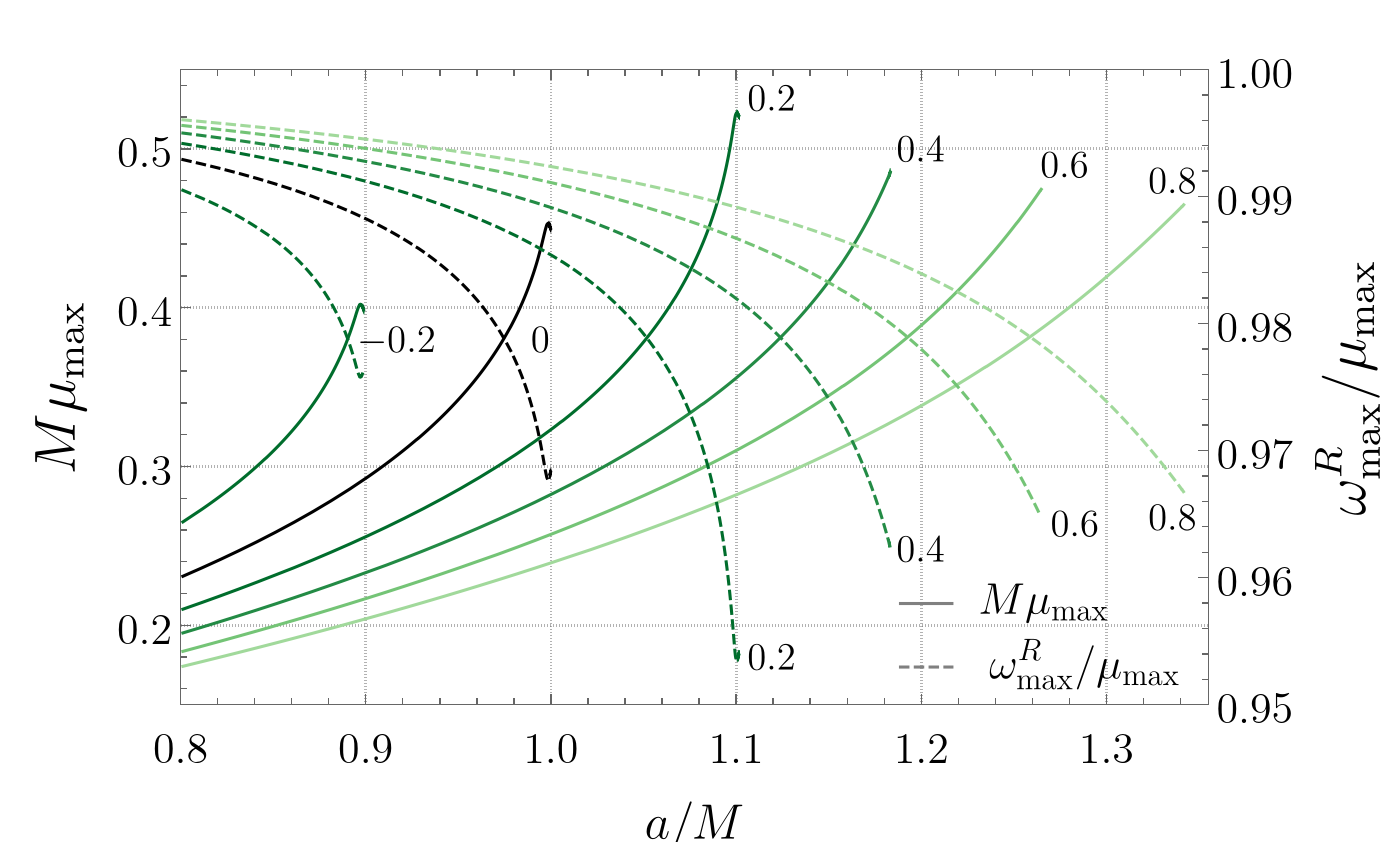} 
     \caption{The quantities $M\omega^{I}_{\mathrm{max}}$ (top panel), $M\mu_{\mathrm{max}}$ (bottom panel, solid lines) and $\omega^{R}_{\mathrm{max}}/\mu_{\mathrm{max}}$  (bottom panel, dashed lines) associated with the $\ell=m=1$ superradiant instabilities as a function of the spin for selected values of deformation (shown next to each curve). The black curves correspond to Kerr black holes.}
  \label{fig:max_ins2} 
\end{figure}

We remark, however, that the general characteristic described above may change when extremality is approached. Such a change of behavior is identified if one refines the parameter space near the red dotted line of Fig.~\ref{fig:horizons} that corresponds to extremal black holes. In fact, for fixed deformation, the maximum instability $\omega^{I}_{\mathrm{max}}$ does not increase monotonically with the spin - it reaches a peak value close to the corresponding extremal spin and then decreases. We illustrate this in the top panel of Fig.~\ref{fig:max_ins2}, where we plot $M\omega^{I}_{\mathrm{max}}$ as a function of the spin for selected deformation parameters. In particular, we observe the concave shape close to the maximal spin for $\eta/M = -0.2$, $\eta/M =0$, and $\eta/M =0.2$. On the other hand, in the case of deformations which do not allow for extremal black holes, namely $\eta/M = 0.4$, $\eta/M =0.6$, and $\eta/M =0.8$, we observe monotonicity. These observations are complemented by the bottom panel of Fig.~\ref{fig:max_ins2}, where we plot $M\mu_{\mathrm{max}}$ and  $\omega^{R}_{\mathrm{max}}/\mu_{\mathrm{max}}$ for the same selection of deformation parameters.

Figs.~\ref{fig:max_ins} and \ref{fig:max_ins2} demonstrate that the superradiant instabilities can be at least one order of magnitude stronger for the deformed Kerr black hole in comparison to the standard Kerr black hole. In fact, for a Kerr black hole the maximum instability reaches the peak value $M\omega^{I}_{\mathrm{max}} \approx 1.72 \times 10^{-7}$ at $a/M \approx 0.997$. The corresponding oscillation frequency and mass are, respectively, $M\omega^{R}_{\mathrm{max}} \approx 0.438$ and $M\mu_{\mathrm{max}} \approx 0.449$. In the darkest region of the left panel in Fig.~\ref{fig:max_ins} we note that the peak value $M\omega^{I}_{\mathrm{max}}$ surpasses $10^{-6}$. Considering the whole parameter space studied in Fig.~\ref{fig:max_ins}, the maximum instability we have found is $M \omega^{I}_{\mathrm{max}} \approx 1.49 \times 10^{-6}$ and occurs at $\left(a/M,\eta/M\right) \approx \left(1.151,0.2913 \right)$, which is very close to the triple point $(2/\sqrt{3},8/27)$. The corresponding oscillation frequency and mass are, respectively, $M\omega^{R}_{\mathrm{max}} \approx 0.582$ and $M\mu_{\mathrm{max}} \approx 0.637$.


\section{Final remarks}
In this work we have used a semi-analytical method to calculate the QNMs and the QBSs of a massive scalar field around a Kerr-like black hole. The deformed Kerr black hole we have considered, described by the metric \eqref{eq:deformed_Kerr_line_element}, is a subclass of the general KRZ parametrization~\cite{KRZ;2016}. Taking advantage of the symmetries of the Kerr black hole which are preserved by the deformations (in particular, the separability of the Klein-Gordon equation and the spherical symmetry of the event horizon), we were able to implement the continued fraction method. Our main goal was to analyze how the parameter $\eta/M$ that quantifies deviations from Kerr, together with the spin parameter $a/M$, influences the eigenfrequencies of a massive scalar field around the black hole. Our results thus complement the analysis of Ref.~\cite{KZ;2016}, according to which deviations from Kerr can produce QNMs which are compatible with the QNMs of a Kerr black hole.

Regarding the stability of the deformed Kerr black hole \eqref{eq:deformed_Kerr_line_element}, we have not found any eigenfrequency with positive imaginary part when $\mu M=0$. We have, nevertheless, demonstrated that zero-damped modes, which are characteristic of extremal Kerr black holes, also exist for deformed Kerr black holes if the deformation parameter satisfies $0 \le \eta/M \le 8/27$. Our findings suggest that the deformed Kerr black hole, like the standard Kerr black hole, is mode stable to massless scalar perturbations. If the field is massive, on the other hand, we have found that superradiant instabilities, as in the case of the standard Kerr black hole, are present. 

A brief investigation of superradiant instabilities in the deformed Kerr metric was performed in Ref.~\cite{Stefano2021} using approximate analytical methods. Our work expands the results of Ref.~\cite{Stefano2021} by analyzing in detail the dependence of such instabilities as a function of the black hole parameters and the mass of the scalar field.  In particular, for each set of  black hole parameters, we were able to determine the mass of the scalar field which maximizes the instability (i.e.~minimizes the growth timescale associated with it). We have observed that, as in the case of the standard Kerr black hole, the peak instability occurs close to (but not at) the extremal spin. Nevertheless, at sufficiently large deformations $\eta/M > 8/27$, for which extremal black holes are not possible, the peak instability increases monotonically with the spin in the range covered by our analysis. Scanning the entire parameter space considered in our work, we find that the growth timescale can be one order of magnitude shorter for a deformed Kerr black hole in comparison to the standard case.                  

We have also provided a detailed investigation of the masses and the oscillation frequencies associated with scalar clouds around the deformed Kerr metric \eqref{eq:deformed_Kerr_line_element} as the black hole parameters vary.  We believe our work can motivate further investigations of hairy black black hole solutions in gravity theories beyond General Relativity. For instance, it would be interesting to study the solutions of the Klein-Gordon field coupled to the gravitational field in pseudo-complex General Relativity, where the metric \eqref{eq:deformed_Kerr_line_element} was first discussed. We also believe that our work, together with Refs.~\cite{Stefano2021,KZ;2016}, will provide new insights and motivate further investigation of the astrophysical applications (including, but not limited to, GW observations) of the deformed Kerr black hole \eqref{eq:deformed_Kerr_line_element} and the KRZ parametrization.


\acknowledgments

We are grateful to R.~A.~Mosna and O.~Zhidenko for useful comments. This research was partially financed by the Coordena\c{c}\~ao de Aperfei\c{c}oamento de Pessoal de N\'{i}vel Superior (CAPES, Brazil) - Finance Code 001. M.~R.~acknowledges support from the Conselho Nacional de Desenvolvimento Cient\'{i}fico e Tecnol\'{o}gico (CNPq, Brazil), Grant No. FA 315664/2020-7.


\bibliography{ref}

\end{document}